\documentclass[structabstract]{aa}  
\usepackage{tabularx}                                 
\usepackage{graphicx}
\usepackage[varg]{txfonts}
\usepackage[sort]{natbib}

\begin{document}

   \title{Ram pressure and dusty red galaxies - key factors in the evolution of the multiple cluster system Abell 901/902\thanks{Based on observations with the European Southern Observatory Very Large Telescope (ESO-VLT), observing run ID 384.A-0813.}}

   \author{Benjamin B\"{o}sch\inst{1}\fnmsep\thanks{\email{benjamin.boesch@uibk.ac.at}}
          \and Asmus B\"{o}hm\inst{1}
          \and Christian Wolf\inst{2}
          \and Alfonso Arag\'{o}n-Salamanca\inst{3}
          \and Marco Barden\inst{1}
          \and Meghan E. Gray\inst{3}
          \and Bodo L. Ziegler\inst{4}
          \and Sabine Schindler\inst{1}
          \and Michael Balogh\inst{5}
          }

   \institute{
Institute for Astro- and Particle Physics, University of Innsbruck, Technikerstr. 25/8, A-6020 Innsbruck, Austria
\and Department of Physics, Denys Wilkinson Building, University of Oxford, Keble Road, Oxford OX1 3RH, UK
 \and School of Physics and Astronomy, The University of Nottingham, University Park, Nottingham NG7 2RD, UK
 \and Department of Astronomy, University of Vienna, T\"{u}rkenschanzstr. 17, 1180 Wien, Austria       
 \and Department of Physics, University of Waterloo, Waterloo, ON N2L 3G1, Canada
  }          
   \date{Received 19 March 2012 / Accepted 16 November 2012}

  \abstract
   {We present spectroscopic observations of 182 disk galaxies (96 in the cluster and 86 in the field environment) in the region of the Abell 901/902 multiple cluster system, which is located at a redshift of $z\sim 0.165$. We estimate dynamical parameters of the four subclusters and analyse the kinematics of spiral galaxies, searching for indications of ram-pressure stripping. Furthermore, we focus on dusty red galaxies as a possible intermediate stage in the transformation of field galaxies to lenticulars when falling into the cluster.}
   {We obtained multi-object slit spectroscopy using the VLT instrument VIMOS. We carried out a redshift analysis, determined velocity dispersions using biweight statistics, and detected possible substructures with the Dressler-Shectman test. We exploited rotation curves from emission lines to analyse distortions in the gaseous disk of a galaxy, as well as HST/ACS images to quantify the morphological distortions of the stellar disk.}
   {The presence of substructures and non-Gaussian redshift distributions indicate that the cluster system is dynamically young and not in a virialised state. We find evidence of two important galaxy populations. \textit{Morphologically distorted galaxies} are probably subject to increased tidal interactions. They show pronounced rotation-curve asymmetries at intermediate cluster-centric radii and low rest-frame peculiar velocities. \textit{Morphologically undistorted galaxies} show the strongest rotation-curve asymmetries at high rest-frame velocities and low cluster-centric radii. Supposedly, this group is strongly affected by ram-pressure stripping due to interaction with the intra-cluster medium. Among the morphologically undistorted galaxies, dusty red galaxies have particularly strong rotation-curve asymmetries, suggesting that ram pressure is an important factor in these galaxies. Furthermore, dusty red galaxies have on average a bulge-to-total ratio that is higher by a factor of two than cluster blue-cloud and field galaxies.\\
The fraction of kinematically distorted galaxies is $75\%$ higher in the cluster than in the field environment. This difference mainly stems from morphologically undistorted galaxies, indicating a cluster-specific interaction process that only affects the gas kinematics but not the stellar morphology. Also the ratio between gas and stellar scale length is reduced for cluster galaxies compared to the field sample. Both findings could be explained best by ram-pressure effects.
 }
   {Ram-pressure stripping seems to be an important interaction process in the multiple cluster system A901/902. Dusty red galaxies might be a crucial element in understanding the transformation of field disk galaxies into cluster lenticular galaxies.}

               
\authorrunning{B. B\"{o}sch et al.} 
\titlerunning{Ram pressure and dusty red galaxies - key factors in the evolution of A901/02}

   \maketitle
%
\section{Introduction}

Ever since the pioneering work by \citet{dressler80}, growing evidence has been found for multiple links between galaxy evolution and environment. Early-type galaxies (elliptical and S0 or lenticular galaxies) are predominant in dense regions like cluster cores, whereas late-type galaxies are more abundant in less dense field environments. Correlations have also been found between galaxy density and colour \citep[e.g.][]{blanton05} or star-formation rate \citep[e.g.][]{verdugo08}. Active galactic nuclei (AGN) seem to be more frequent in moderate-density environments \citep{gilmour07,kauffmann03}. \\
Disentangling the relative importance of the internal and external physical mechanisms responsible for these relations is challenging. Galaxies in cluster environments may differ from field galaxies owing to higher initial densities, leading to earlier collapse. This could mean a difference in evolution due to their ``nature". But also the influence of ``nurture" can be important. Both Galaxy-galaxy interactions, such as tidal interactions, major/minor mergers, or harassment \citep{moore96}, and galaxy-cluster interactions, such as ram-pressure stripping \citep{gunn72,kronberger08} due to the ICM (Intra-cluster medium), halo truncation \citep{balogh00,bekki01}, or gas compression \citep{tonnesen09,byrd90} can play crucial roles in the evolution of galaxies. To decouple the different effects of environment on galaxy evolution, it is necessary to analyse a variety of galaxy properties, such as kinematics, spectral energy distributions (SEDs), morphologies, stellar masses, or star-formation rates (SFRs) over a wide range of environments.\\
 Some properties of high-density environments remain puzzling. One of these is the large increase in the fraction of lenticular galaxies in local galaxy clusters compared to higher redshifts, accompanied by a proportional decrease in the spiral fraction \citep{dressler97}. A possible scenario would be that spiral galaxies falling from the field into a cluster's gravitational potential are transformed into S0 galaxies. Ram-pressure stripping of the infalling galaxies' gas due to the ICM or starvation due to the cluster potential \citep[e.g.][]{balogh00} might be important factors in this transformation and responsible for the truncation of star formation. Besides this disk fading there would need to be some mechanism that causes the bulges in spirals to grow, since bulges in S0s are more luminous than those in spirals \citep[e.g.][]{christlein04}. Galaxy harassment \citep{moore96} or tidal interactions could be such mechanisms and hence might contribute to the transformation process. The dominant evolutionary path is still unknown.\\
	In their analysis of 10 intermediate-redshift clusters ($z\sim 0.4-0.6$), \citet{poggianti99} find a suppression of star formation in cluster members. Furthermore they propose a candidate for an intermediate phase during the spiral-S0 transformation. They identify a significant fraction of dusty starburst galaxies (strong Balmer absorption, modest [\ion{O}{II}] emission), probably falling in from the surrounding field. The descendants of these galaxies could be poststarburst (k+a) galaxies and passive spirals. They suggest that the resulting passive disk populations might be the progenitors of S0 galaxies \citep[see also][]{jones00}. In agreement with this hypothesis,\citet{kocevski11} find an obscured starburst population in cluster environments more frequently than in the field  using Spitzer-$24\mu m$ imaging. They propose that, due to a mix of mergers and harassment, a higher fraction of cluster and group members are experiencing this nuclear starburst activity compared to field galaxies, and might be transformed into lenticulars with strong bulges.\\
		Another intermediate phase during this process could be the class of \textit{``dusty red spirals"} \citep{wolf03}. Having star-formation rates four times lower at fixed mass compared to blue spirals \citep{wolf09} and showing similar extinction levels, dusty red galaxies can be understood as the low specific-SFR tail of the blue cloud. They show only weak spiral structures and appear predominantly in the stellar mass range of $\log (M_{\ast}/M_{\odot}) = [10,11]$, where they represent over half of the star-forming galaxies in the A901/902 cluster system \citep{wolf09}. Such a rich dusty red population indicates that if star-formation quenching does indeed happen, it must be a slow process and that any morphological transformation is delayed. At $\log (M_{\ast}/M_{\odot}) <10$ dusty reds are rare, suggesting a fast transformation in the low mass regime.\\
The STAGES (Space Telescope Abell 901/902 Galaxy Evolution Survey) project \citep{gray09} and its large amount of data (see Section 2) provides information including galaxy morphologies, SEDs, SFRs and stellar masses for a specific region on the sky -- the multiple cluster system Abell 901/902. This system might still be in the process of formation and not yet be virialised. It is therefore an interesting laboratory to investigate the interplay between galaxy evolution and environment. \citet{gallazzi09} demonstrate the general suppression of star formation at higher galaxy number densities in the STAGES system, and find that a high fraction of the remaining star formation in the cluster is either obscured or dominated by old stellar populations. Therefore dusty star-forming galaxies are predominant at intermediate densities. But there are also examples of a lack of dependence of galaxy properties on environment: \citet{maltby12} find for STAGES spirals no dependence of the stellar distribution in the outer stellar disk on galaxy environment and conclude that differences in this distribution are related to internal mechanisms or minor mergers. \\
In this work we exploit spatially resolved spectra from the VLT (Very Large Telescope) instrument VIMOS (VIsible MultiObject Spectrograph), complemented with already existing data from the STAGES survey \citep{gray09}. The paper is organised as follows. Section 2 gives a short introduction into the STAGES project, reports on the target selection and observation of our VLT-VIMOS spectroscopy and describes the data reduction process. Section 3 presents the redshift analysis and gives an overview of the subcluster properties. Section 4 represents the main part and contains the kinematic analysis exploiting rotation curve asymmetries. The last subsection deals with the prominent role of dusty red galaxies. A summary is provided in Section 5.\\
  Throughout this paper, we assume a cosmology with $\Omega_{m}=0.3$, $\Omega_{\Lambda}=0.7$ and $H_{0}=70 \mbox{kms}^{-1} \mbox{Mpc}^{-1}$. 


\section{Observations and data reduction}
\subsection{The STAGES data set}
This work utilises data from the STAGES project. STAGES is a multi-wavelength survey that covers a wide range of galaxy luminosities and galaxy densities.\\
The structure under scrutiny is the multiple cluster system Abell 901/902 at $ z \sim 0.165 $, which comprises 4 subclusters and has been the subject of V-band (F606W) Hubble Space Telescope/Advanced Camera for Surveys (HST/ACS) imaging, covering a $0\fdg5\times0\fdg5$ ($\sim5\times5 \textrm{Mpc}^{2}$) area of the multi-cluster system. Additionally 17-band COMBO-17 (Classifying Objects by Medium-Band Observations), Spitzer 24 $\mu m$, XMM-Newton X-ray data \citep{gilmour07} and gravitational lensing maps \citep{heymans08} are available.  Photometric redshifts with errors $ \delta_{z}/(1+z) \sim 0.02 $ up to $R = 24$, spectral energy distributions \citep{wolf03} and stellar masses \citep{borch06} have been derived from the COMBO-17 survey as well as star-formation rates from the COMBO-17 UV and Spitzer 24 $\mu m$ data \citep{bell07}. Data from the Galaxy Evolution Explorer (GALEX), 2 degree Field (2dF) spectrograph, and Giant Meterwave Radio Telescope (GMRT) are also available. An overview of the publicly available data is given in \cite{gray09}.
\subsection{VLT observations}
The spectroscopic observations of the Abell cluster system A901/02 with the VLT instrument VIMOS in MOS (Multiobject spectroscopy) mode were completed between February 8 and March 10, 2010 (ESO-ID 384.A-0813, P.I. A. B\"{o}hm). Because of the large field of view, the optical path of VIMOS is split into four channels (also called quadrants). The field of view of these four channels is $7' \times 8'$ each, separated by $2'$ gaps. The CCD area of one quadrant is $2048 \textrm { (spatial axis}) \times 4096  \textrm{ (wavelength axis)}$ pixels. The instrument provides an image scale of $0\farcs205/$pixel. The grism of our choice was the high-resolution grism \textit{HR-blue} to obtain spatially resolved spectra, including the [\ion{O}{II}] emission line doublet at 3726/29 $\AA$. Depending on the slit position on the CCD, a different $\sim 2050\AA $-range of the spectral range ($[3700\AA,6740\AA]$) is seen on the CCD. The HR-blue grism has an average dispersion of $0.51\AA /$pixel and a spectral resolution of $R\sim 2000$ at a slit width of  $1\farcs2$. The seeing conditions covered the range $0.53''\leq \mathrm{FWHM} \leq 1.14''$ according to the Differential Image Motion Monitor (DIMM).\\
Four VIMOS pointings (with $\sim 1'$ overlap) were used to cover the 1/4 square degree A901/902 field across the infall regions to the cluster centres. Two exposures of 1800 seconds each were taken for each pointing.\\
The spectroscopic targets from the STAGES catalogue were selected to have stellar mass $M_{*}>10^{9} M_{\odot}$, absolute magnitude $M_{B}<-18$, a star-forming SED (selection criterion in the public catalogue \citet{stagescat09}: $\mathrm{sed\_type} \ge 2$), photometric redshift $0.1<z_{\mathrm{phot}}<0.26$, and a visually confirmed disk component on the ACS images with an inclination angle $i>30°$. This selection criterion resulted in $\sim 320$ cluster and $\sim 160 $ field candidates.\\
Using MOS slits with tilt angles up to $\pm 45^{\circ}$, aligned along the apparent major axis of the galaxies, a position angle interval of $90^{\circ}$ was covered. The manual slit placement was performed using the ESO software VMMPS (VIMOS Mask Preparation Software). Eventually, $238$ slits were successfully placed. Some objects were included in two or more set-ups due to lack of alternate targets within the given sky area. These galaxies will later be used for consistency checks of our data reduction and analysis. Hence, spectra of $215$ different galaxies were obtained.
\subsection{Spectroscopic data reduction}
The basic reduction steps are performed using the ESO-VIMOS pipeline (version 2.2.3) together with Gasgano (version 2.3.0), a Java-based data file organiser developed and maintained by ESO. For additional reduction steps and further processing we use the software package MIDAS\footnote[1]{Midas, the Munich Image Data Analysis System is developed and maintained by the European Southern Observatory (ESO)}. \\
First, we create a combined  master-bias with the pipeline recipe \textit{vmbias}. We apply the master-bias in the reduction of the flat field, arc lamp, and scientific exposures. We process the available flat field exposures with the recipe \textit{vmspflat}. This produces a normalised master-screen-flat and a non-normalised flat field, called combined-screen-flat, which we use to determine the spectral curvature model. However, we do not utilise the master flat in further reduction steps, since ``the HR-blue grism data show not yet understood reflections, particularly evident in the flat-fields and arc-lamp frames" (excerpt of ESO readme for MOS users).\\
In a next step the recipe \textit{vmspcaldisp} processes the arc lamp exposure and calculates the spectral distortion model, mentioned above. Additionally, a cleaning of bad pixels and cosmic rays is performed. An output table, called \textit{extraction table}, contains the coefficients of the wavelength calibration. In most cases we could provide an \textit{extraction table} generated from previous runs to the recipe as input, in order to iterate the modelling of the spectral distortions. The wavelength calibration spectra with slits in the upper third of the CCD result in a low number of arc lamp lines. We tackle this issue by adding Argon-lines to the default reference catalogue.\\
To achieve a satisfying wavelength calibration even for slits with a large tilt angle, we change a few crucial parameters from their default values. We increase the size of the search window (keyword \textit{vimos.Parameters.extraction.window}) around the expected arc line position to 30 pixels. We lower the threshold for detecting a spectral feature (\textit{vimos.Parameters.line.ident.level}) in order not to miss weak arc lines. Furthermore, we change the order of the polynomial (\textit{vimos.Parameters.slit.order}) for wavelength solution modelling along the spatial axis within each slit from 0 to 1.\\
Finally, the recipe \textit{vmmosobsstare} reduces the science frames including cosmic cleaning, sky subtraction, wavelength calibration, slit rectification and 2D-spectra extraction. \\
An additional cosmic cleaning, a co-addition of the two exposures, and a 1D-object-spectra extraction is realised subsequently in MIDAS.   
\subsection{Spectroscopic redshifts}
\begin{figure}[h]
   \centering
   \includegraphics[angle=0,width=\columnwidth]{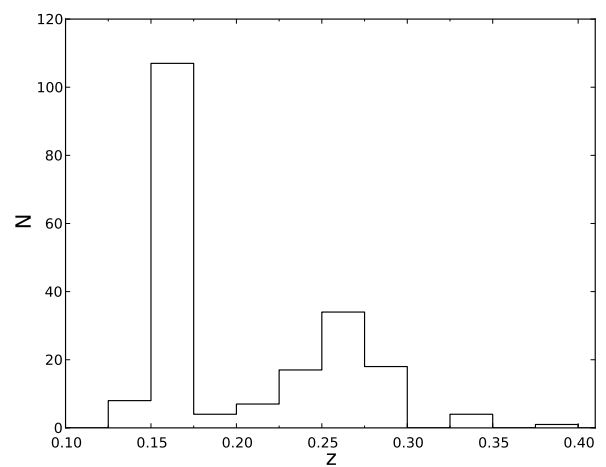}
      \caption{Redshift histogram of the 200 galaxies of our sample having determined redshifts, peaking at the cluster centre at $z\sim0.165$.
              }
         \label{fig:zhist}
   \end{figure}

For 188 of 215 galaxies we could measure spectroscopic galaxy redshifts using emission lines. For another 12 galaxies we could determine the redshift using absorption lines. These galaxies, though selected as star forming, show no detectable emission lines in their spectra. For 15 objects ($\sim 7\%$) no redshift measurement is possible due to the signal-to-noise ratio being too low. Note that none of these are primary targets but are intstead ``fill-up" objects. We show a redshift histogram of all 200 galaxies in Figure \ref{fig:zhist}, clearly peaking at the cluster centre at $z\sim0.165$. The ``bump" at $z\sim0.26$ is probably a volume effect.\\
Errors for galaxy redshifts can be estimated using the spectra of galaxies that have been observed in more than one mask \citep{mj08}. For 16 galaxies we obtained two redshifts, for two objects three redshifts and for one object we even have four redshifts. For each galaxy and redshift, we calculate the difference between the redshift and the mean of the redshifts available for a certain object. To compensate for not calculating the differences with respect to the mean of the underlying distribution we multiply these differences by a scaling factor ($>1$) inferred from Monte-Carlo simulations by \citet{mj08}. \\
We then calculated the biweight estimator of scale \citep{beers90} of the 42 differences to obtain a robust estimator for the standard deviation of the error distribution. It yields $\delta_{z} = 0.00019$ as the estimate of the typical redshift error. For the cluster redshift $z\sim0.165$, this error corresponds to $\delta_{z}/(1+z) \sim 0.00016 $ or  $47 \mathrm{km/s}$ in rest-frame velocity.

\section{Properties of subclusters}
\subsection{Subcluster redshifts and velocity dispersions}
\begin{figure}[h]
   \centering
   \includegraphics[angle=0,width=\columnwidth]{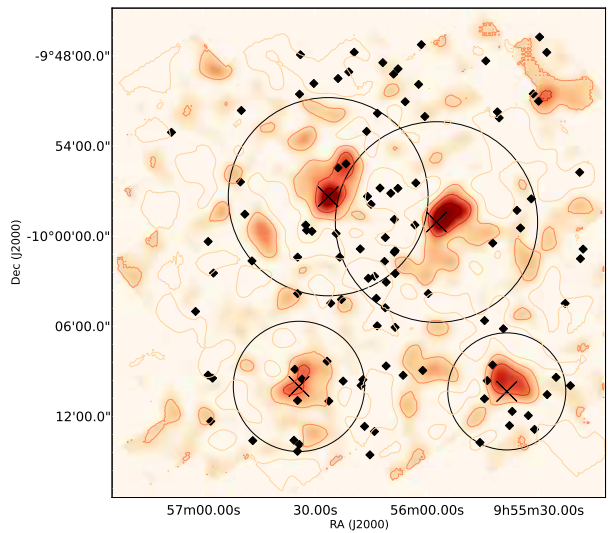}
      \caption{The dark matter reconstruction of the multiple cluster system A901/902 in orange-red colour scale derived from weak-lensing analysis \citep{heymans08}. Cluster members as a result of the biweight analysis (Section 3.1) are indicated with black diamonds. The four big crosses indicate the subcluster centres of A901a, A901b, A902 and SW group (from top left to bottom right), respectivel.y The radii of the 4 big circles are equal to the virial radii $r_{200}$ H08 (see Table \ref{table:mvir2}).    
              }
         \label{fig:vr}
   \end{figure}

\begin{figure}[h]
   \centering
   \includegraphics[angle=0,width=\columnwidth]{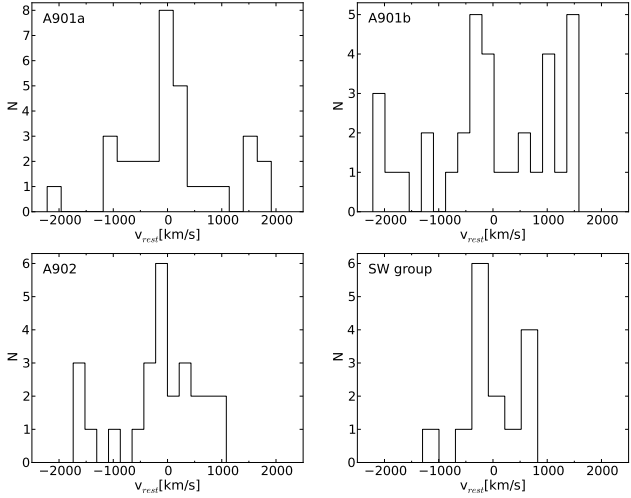}
      \caption{Histograms of peculiar velocities in the cluster rest-frame, for the 4 subclusters. 
              }
         \label{fig:histscl}
\end{figure}

Measurements of velocity dispersion and mean redshift provide a way to estimate the mass of a cluster, and to decide about cluster membership for each galaxy. These calculations should be made using a robust statistic. We here adopt the biweight statistic \citep{beers90}.\\ 
Previous weak lensing analysis of the Abell 901/902 cluster system \citep{heymans08} indicated four subclusters in the dark matter distribution, namely A901a, A901b, A902 and the SW-group (see Figure \ref{fig:vr}). We use the coordinates of the four dark matter peaks as subcluster centres in our following analysis.\\
Furthermore, we apply the biweight estimators of location and scale to determine the mean redshift $z_{\mathrm{scl}}$ as well as the velocity dispersion $\sigma_{\mathrm{scl}}$ of each subcluster. For datasets up to $\sim50$ galaxies the biweight statistic is superior to other estimators and insensitive to outliers \citep{beers90}. \\
We select member candidates for each subcluster as follows. First we apply a common filter in redshift space ($0.15 < z_{i}<0.18$), which corresponds to a $\sim4500 \mathrm{km/s}$ interval around the cluster redshift $z_{\mathrm{scl}}$. This already excludes most background and foreground field galaxies. We then assign each selected galaxy to the nearest subcluster, in terms of projected distance to the subcluster centre on the sky.\\
 For further analysis we transform the galaxy redshifts $z_{i}$ to recession velocities $v_{i}$ using the relativistic formula:
  \begin{equation}
      v_{i} = c\cdot\frac{(1+z_{i})^{2}-1}{(1+z_{i})^{2}+1,} \quad\quad \mbox{where} c:\mbox{ speed of light.}
      \label{eq:relv}
   \end{equation}
The biweight location estimator of these velocities $v_{i}$ gives an initial approximation of the redshift of each subcluster, $z_{scl}$. To obtain the final rest-frame velocity dispersion $\sigma_{\mathrm{scl}}$, we transform all velocities $v_{i}$ of a particular subcluster into its rest-frame:
  \begin{equation}
      v_{i_{\mathrm{rest}}} = \frac{v-v_{\mathrm{scl}}}{1+z_{\mathrm{scl}}}.
      \label{eq:restv}
   \end{equation}
With these rest-frame velocities $v_{i_{\mathrm{rest}}}$ as input data, the velocity dispersion $\sigma_{\mathrm{scl}}$ can be calculated as the biweight estimator of scale, whereas the biweight estimator of location refines the initial guess of the subcluster redshift $z_{\mathrm{scl}}$. \\
For all galaxies in the interval $[-3\sigma_{\mathrm{scl}},+3\sigma_{\mathrm{scl}}]$ the biweight statistic is iteratively calculated in a loop until convergence, i.e. until all input galaxies fall into the $3\sigma$-interval. Accordingly, we define those galaxies to be \textit{cluster members} of the respective subcluster. \\
By generating bootstrap samples from the final set of galaxy velocities, $v_{i_{\mathrm{rest}}}$ \citep{beers90}, we perform Monte Carlo simulations (2000 replications) to estimate the asymmetric $68\%$ error bars. For each bootstrap sample we determine the biweight estimator of scale (velocity dispersion $\sigma_{\mathrm{scl}}$) and of location (subcluster redshift $z_{\mathrm{scl}}$). All these values combined simulate the underlying distributions. In a first approximation the mean and standard deviation of each distribution give confidence intervals. Since the distributions of the bootstrap scale estimators are usually not Gaussian, another biweight is performed. From the confidence intervals we infer the error bars of the biweight estimators of the original sample, by calculating the difference to the interval limits. \\
  We employ this procedure to determine the velocity dispersion and mean redshift of all four subclusters as well as their asymmetric error bars. The results are summarised in Table \ref{table:vdisp}. The values for the velocity dispersion agree with those derived by Gray et. al (in prep.) from 2dF spectra. Figure \ref{fig:histscl} shows histograms of the peculiar velocities in subcluster restframe $v_{i_{\mathrm{rest}}}$ of the final member galaxies calculated for the given value of $z_{\mathrm{scl}}$. Notably, the subcluster A901b has many objects in the wings of the distribution, which may really be part of another subcluster. We will discuss this matter quantitatively in Subsection 3.2. \\

\begin{table}    
\centering                          
\begin{tabular}{c c c c}        
\hline\hline                 
Subcluster & Member& $z_{\mathrm{scl}}$ & $\sigma_{\mathrm{scl}}$ \\&galaxies $N$&&[km/s]\\    
\hline            
\noalign{\smallskip}   
   A901a & 31 & $0.1642^{+0.0006}_{-0.0006}$ & $901^{+150}_{-203}$ \\      
\noalign{\smallskip}  
   A901b & 34 & $0.1628^{+0.0004}_{-0.0004}$ & $1175^{+100}_{-130}$ \\
\noalign{\smallskip}    
   A902 & 26 & $0.1665^{+0.0006}_{-0.0006}$ & $792^{+150}_{-203}$ \\
\noalign{\smallskip}    
   SW group & 15 & $0.1689^{+0.0004}_{-0.0004}$ & $552^{+97}_{-158}$ \\
\noalign{\smallskip}  
\hline                                   
\end{tabular}
\caption{Velocity dispersion and mean redshift of the 4 subclusters. Additionally the number $N$ of $3\sigma$-member galaxies of each subcluster is given.}             
\label{table:vdisp} 
\end{table}

\subsection{Substructure}
\begin{figure*}
   \centering
   \includegraphics[angle=0,width=\textwidth]{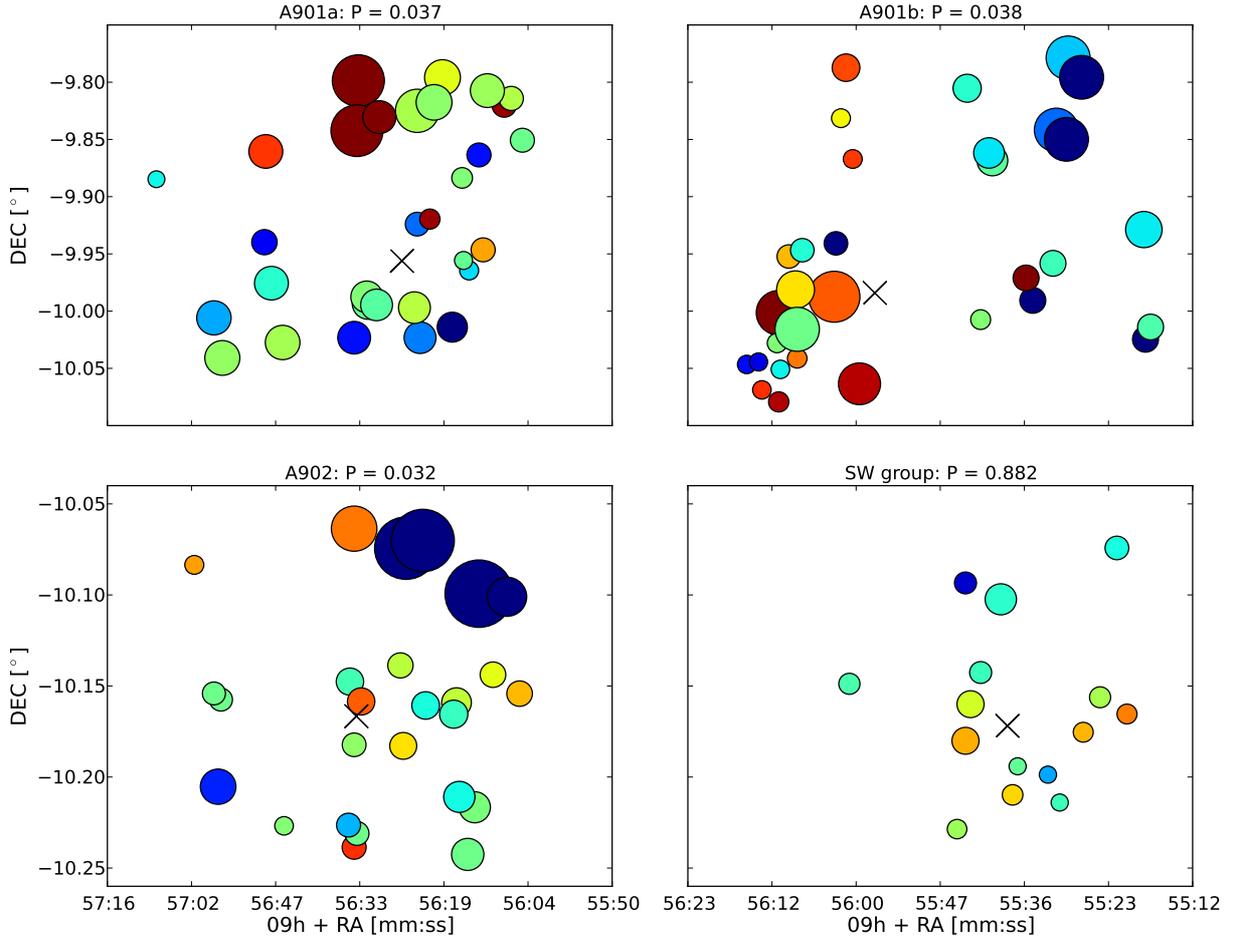}
      \caption{Results from the Dressler-Shectman (DS) test. The plots show the spatial location of the cluster member galaxies. The radii of the plotted circles are proportional to $e^{\delta}$, where $\delta$ is the DS measurement of the local deviation from the global velocity dispersion and mean recessional velocity, i.e. larger symbols correspond to a higher significance of residing in a substructure. The colours indicate the rest-frame velocity relative to the subcluster centre, ranging from red (receding; 1500km/s) to blue (approaching; -1500km/s). The black crosses indicate the subcluster centres. $P$ is the probability of there being no substructure in the dataset. Except for the SW-group, each of the other 3 subclusters shows a high significance for substructure. The significant substructures and non-Gaussian redshift distributions indicate that the cluster system is dynamically young and not in a virialised state.
}
         \label{fig:dstest}
   \end{figure*}

To quantify the existence of substructure in galaxy clusters many statistical tests have been devised. \citet{ds88} developed a method to estimate the significance and location of cluster substructure using galaxy redshifts and projected sky positions. We implement this three-dimensional (velocity-position) test in our analysis and apply it to each subcluster. \\
Starting from a list of (sub)cluster members with measured positions (ra, dec) and velocities, we determine for each galaxy the local mean velocity $\bar v_{\mathrm{local}} $ and velocity dispersion $\sigma_{\mathrm{local}}$ from the sample of $N_{\mathrm{nn}}$ nearest neighbours on the sky. Following \citet{pinkney96} we used $N_{\mathrm{nn}}=\sqrt{N}$ for the number of nearest neighbours. This maximises the sensitivity to significant substructures while reducing the sensitivity to Poisson noise. Next, we compare these local values to the global mean velocity $\bar v$ and velocity dispersion $\sigma$ of the whole subcluster (see Table \ref{table:vdisp}). A measure for the deviation from the global values can then be defined for each galaxy as
\begin{equation}
      \delta^{2} = \frac{ N_{\mathrm{nn}} +1} {\sigma^2}\cdot\left[\left(\bar v_{\mathrm{local}}-\bar v\right)^{2}+\left(\sigma_{\mathrm{local}}-\sigma\right)^{2}\right].
      \label{eq:dstest}
 \end{equation}
 Eventually, we compute the cumulative deviation $\Delta=\Sigma\delta$, which serves as a statistic for quantifying the substructure. By comparing the $\Delta$ statistic with a set of 1000 Monte Carlo simulations by randomly reshuffling the velocities of the subcluster members, we can quantify the statistical significance of a substructure by the fraction $P$ of simulations that result in a higher $\Delta$ value than the observed one \citep{ds88}. A small value of $P$ corresponds to a high significance. The criterion $P\leq0.1$ is commonly used as an indicator for substructures \citep[e.g.][]{popesso07}.\\
 We visualise the possible position of substructures in Figure \ref{fig:dstest}: markers whose size is proportional to $e^{\delta}$ depict the distribution of member galaxies on the sky, thus quantifying the local deviation from the global kinematics. Additionally, the symbols are coloured according to the subcluster rest-frame peculiar velocity of a galaxy. Some large circles in a given area indicate a correlated spatial and kinematic variation and may locate kinematically cooler systems within the cluster (e.g. subgroups) as well as flows of infalling galaxies. It is noteworthy, that the $\Delta$ statistic is  insensitive in situations where substructures are superimposed. It relies on some spatial displacement of the centroids.\\
 Except for the SW-group, each of the other 3 subclusters shows significant substructures. For A902 (lower left) a clump of galaxies (shown in blue) indicates a substructure that might be associated with the more blueshifted subcluster A901a. A901b shows the strongest subclustering, in terms of having a high probability for substructure in combination with the clearly non-Gaussian galaxy velocity distribution (see Figure \ref{fig:histscl}). Nine member galaxies of A901b have a delta-statistic $\delta>2$. In comparison A901a has four, A902 has five and the SW group no such member galaxies. Accordingly, \citet{heymans08} find for A901a a significant substructure in the dark matter distribution, associated with the infalling X-ray group A901$\mathrm{\alpha}$.\\The presence of substructures and non-Gaussian redshift distributions (see Figure \ref{fig:histscl}) underline that the cluster system is dynamically young and not in a virialised state, which would be characterised by a Gaussian galaxy velocity distribution \citep[e.g.][]{nakamura00, merrall03} and, as indicated by N-body simulations, by a low mass fraction included in substructures \citep[e.g.][]{shaw06}. The presence of such substructures can lead to an overestimation of dynamical parameters, especially when applying the virial theorem (see next subsection).

\subsection{Dynamical parameters of the subclusters}
Estimating dynamical parameters of a cluster is not an easy task and various methods are available.  Applying the virial theorem to positions and velocities of cluster member galaxies is the standard method for estimating the mass of a self-gravitating system. This approach assumes that the system under scrutiny is in dynamical equilibrium and that the galaxies in a cluster trace the mass, which means that the number density is proportional to the mass density. The virial mass estimator is derived from the Jeans equation and relates the total kinetic energy to the potential energy ($2T+U=0$). It is given by \citep[see e.g.][]{carlberg96}: 
\begin{equation}
      M_{v} = \frac{3\sigma^{2}} {G}\frac{\pi R_{\mathrm{pv}}}{2}=\frac{\sigma_{3D}^{2}\cdot r_v}{G},
      \label{eq:virmass}
 \end{equation}
where the three-dimensional velocity dispersion $\sigma_{3D}$ and the three dimensional virial radius $r_v$ can be calculated from observables using the line-of-sight velocity dispersion $\sigma$ (see Section 6) and the projected virial radius $R_{pv}$. For more details see \citet{carlberg96}. \\
One interpretation is that the radius of virialisation corresponds to the radius where the internal material is virialised, but the external material is still plunging into the cluster. Some simulations indicate that this happens at  $r_{200}$, the radius where the mean interior density is $200\rho_c$ \citep{white01}. Accordingly, a widely used mass estimator is $M_{200}$: the mass within $r_{200}$. \\
Note that the masses calculated using Equation \ref{eq:virmass} depend on the radial extent of the sample, which in turn is connected to the mean interior density $\overline{\rho}(r_v)$ inside the virial radius. The mean density $\overline{\rho}$ is scaled to the critical density $\rho_{c}(z)=3H^{2}(z)/8\pi G$, the density for a flat universe at an epoch represented by redshift $z$.\\
To determine these values we follow \citet{carlberg97} and first correct $M_v$ for the surface pressure term ($2T+U=0$ is replaced by $2T+U=3PV$, accounting for not integrating the Jeans equation to infinity), which is the external pressure from matter outside the virialised region \citep[see e.g.][for details]{the86, girardi98}, by employing a typical reduction at $r_v$ of $20\%$ \citep{rines03, castellano11}. Then, by assuming a singular isothermal sphere model for the density-radius relation we extrapolate $r_v$ to $r_{200}$: 
\begin{equation}
       r_{200} =r_{v}\left[\frac{\overline{\rho}(r_v)}{200\rho_c} \right]^{1/2}. 
      \label{eq:r200}
\end{equation}
Subsequently, one can determine the corresponding mass $M_{200}$.\\
Using N-body simulations, \citet{perea90a} show that the virial mass estimator is superior to other two-mass estimators (median mass estimator, projected mass estimator) and is less sensitive to anisotropies or subclustering. However, it is affected by the presence of interlopers, i.e. unbound and non-virialised galaxies, or the existence of a mass spectrum. Each factor can introduce an overestimation of the cluster mass by factors as large as two to four. Thus, the calculated masses have to be considered as a upper limits.\\
Another bias might be introduced by the presence of a velocity-morphology segregation. For many clusters, the star-forming late-type population with blue colours shows a larger velocity dispersion than the red and passive early-type population \citep[e.g.][]{carlberg97, adami98}. For example, in the Coma cluster, factors up to $\sim1.4$ between the velocity dispersion of early- and late-type galaxies are measured \citep{adami98}. \citet{goto05} suggests that evolved massive galaxies might have reduced their velocity by dynamical friction between the cluster members through less vigorous tidal interactions (e.g. galaxy harassment). Since our sample mainly consists of galaxies with a star-forming SED, the measured velocity dispersion might introduce another overestimation of the cluster mass.\\
Many approaches to account for the interloper-problem are discussed in literature. The above calculation of the virial mass (Equation \ref{eq:virmass}) is an idealised estimate, for the case of perfect identification of virialised cluster members. Interloper galaxies lie close to the cluster in projected distance and in redshift but are in fact members of a different halo or substructure. Restricting samples to elliptical galaxies, which unfortunately is not possible in our case, can reduce but not alleviate the problem completely \citep{white10}. \\
\citet{wojtak07} discuss different methods of interloper treatment and test them in detail using the results of cosmological N-body simulations.They find that the methods of \citet{denhartog96} and \citet{perea90a} are the most efficient ones and achieve an identification rate of gravitationally unbound galaxies of $60-70\%$ at negligible rates of false identifications. Here we adopt the method of \citet{perea90a}, which relies on the iterative removal of galaxies whose absence in the sample causes the biggest change in the virial and projected mass estimator.\\
Using only the virialised galaxies identified in this way, we calculate the dynamical parameters of A901/902 (see Table \ref{table:mvir2}). Also shown are the corresponding parameters derived from the weak-lensing analysis \citep[][two-halo model]{heymans08}. Since the authors there define the virial radius $r_{200H}$ as the radius where the mass density of the halo is equal to 200-times the critical matter density $\Omega_m(z)\rho_c(z)$ at the redshift of the cluster, we first have to extrapolate the mass $M_{200}$ to our definition of $r_{200}$ (Equation \ref{eq:r200}). To get an estimation of the velocity dispersion of each subcluster, we then use the $M_{200}$-$\sigma$ relation found by \citet{biviano06}.\\
Although the applied interloper removal algorithm does lower the calculated virial masses, they are partly still significantly higher than the values derived from weak-lensing analysis. This overestimation is a further indication of the non-virialised state of the subclusters and the high significance for substructures obtained using the DS-test. Hence, the dynamical parameters of the dark matter analysis are used hereafter.\\
We emphasise that the results of the following section would not change significantly if the virial theorem based dynamical parameters are used.\\   
Cluster membership is adopted from Section 3.1, since galaxies residing in the neighbourhood of a cluster centre in terms of projected position and velocity can be subject to cluster-specific interactions, even though they might be in a non-virialised state.
\begin{table*}    
\centering                          
\begin{tabular}{c c c c c c c c c c c}        
\hline\hline                 
Subcluster & \# & $\sigma_{\mathrm{scl}}$ &$M_{v}$ & $R_{\mathrm{pv}}$ & $\overline{\rho}(r_v)$ & $r_{200}$ & $M_{200}$ & $\sigma_{\mathrm{scl}}$ H08 & $r_{200}$ H08 & $M_{200}$ H08 \\& galaxies & [km/s] & $(10^{14} \; M_{\odot})$ & (kpc) & $(\rho_c^{-1})$&(kpc) & $(10^{14} \; M_{\odot})$ & [km/s] & (kpc) & $(10^{14} \; M_{\odot})$ \\
\hline            
\noalign{\smallskip}   
   A901a & 31 & $901^{+150}_{-203}$ & $18.4^{+5.4}_{-11.4}$ & $2067^{+159}_{-376}$ & $64^{+40}_{-39}$ & $1842^{+430}_{-604}$ & $8.3^{+3.8}_{-6.2}$ &$974^{+85}_{-85}$ & $1141^{+100}_{-99}$ & $2.0^{+0.6}_{-0.5}$\\
\noalign{\smallskip}  
   A901b & 30 & $954^{+83}_{-123}$ & $20.8^{+6.2}_{-9.7}$ & $2090^{+563}_{-707}$ & $70^{+48}_{-44}$ & $1950^{+162}_{-259}$ & $9.8^{+2.4}_{-3.7}$ &$987^{+74}_{-87}$ & $1155^{+86}_{-102}$ & $2.1^{+0.5}_{-0.5}$\\
\noalign{\smallskip}    
   A902 & 21 & $471^{+56}_{-75}$ & $3.6^{+2.9}_{-3.3}$ & $1474^{+215}_{-300}$ & $34^{+23}_{-18}$ & $960^{+163}_{-303}$ & $1.2^{+0.4}_{-0.8}$ &$640^{+114}_{-105}$ & $750^{+134}_{-123}$ & $0.6^{+0.4}_{-0.2}$\\
\noalign{\smallskip}    
   SW group & 14 & $460^{+45}_{-84}$ & $3.1^{+0.8}_{-2.3}$ & $1344^{+101}_{-395}$ & $39^{+38}_{-34}$ & $938^{+119}_{-219}$ & $1.1^{+0.3}_{-0.7}$ &$576^{+117}_{-98}$ & $674^{+138}_{-115}$ & $0.4^{+0.3}_{-0.2}$\\
\noalign{\smallskip}  
\hline   
\noalign{\smallskip}                              
\end{tabular}
\caption{Dynamical parameters for the 4 subclusters using only ``virialised"-cluster galaxies. The errors are computed via bootstrap simulations (2000 replications). $R_{\mathrm{pv}}$ denotes the projected virial radius, $M_v$ is the virial mass inside the 3D-virial radius $r_v=\frac{\pi}{2} R_{\mathrm{pv}}$. $\overline{\rho}(r_v)$ is the mean interior density inside the virial radius. $r_{200}$ is the radius where the mean interior density is $200\rho_c$. $M_{200}$ is the corresponding mass. As a comparison the corresponding values derived from weak-lensing analysis (H08) following \citet{heymans08} and \citet{biviano06}.}             
\label{table:mvir2} 
\end{table*}

\section{Kinematic analysis of galaxies}
In this section we analyse the kinematics of our galaxy sample. We derive spatially resolved rotation curves from emission lines. Since these rotation curves trace the gaseous disk of a galaxy, asymmetries in such a curve can be a measure for kinematic distortions due to environmental effects like ram-pressure stripping, mergers, galaxy harassment or tidal interactions with the cluster potential. \\
\citet{kronberger08} find in combined N-body/hydrodynamic simulations that the collision-less stellar disk is not affected by ram pressure and only the gaseous component shows disturbed kinematics. Asymmetries in the gas kinematics combined with a lack of distortions in the stellar disk could therefore be a strong indicator for ram-pressure stripping. Furthermore \citet{kronberger08} show that the effects of ram-pressure stripping depend on the orientation of a disk galaxy's plane with respect to its motion through the ICM. Characteristic features are a mismatch between the kinematic and the luminous centre, declining outer parts of the rotation curve or even distortions in the inner parts. However, ram pressure was found not to affect the stellar disk of a galaxy significantly.\\
In this context we want to take a closer look at the population of ``dusty red" galaxies first introduced by \citet{wolf05} and investigate to what extent they differ from ``blue cloud" galaxies.

\subsection{Rotation-curve extraction}
\begin{figure}[h]
   \centering
   \includegraphics[angle=0,width=\columnwidth]{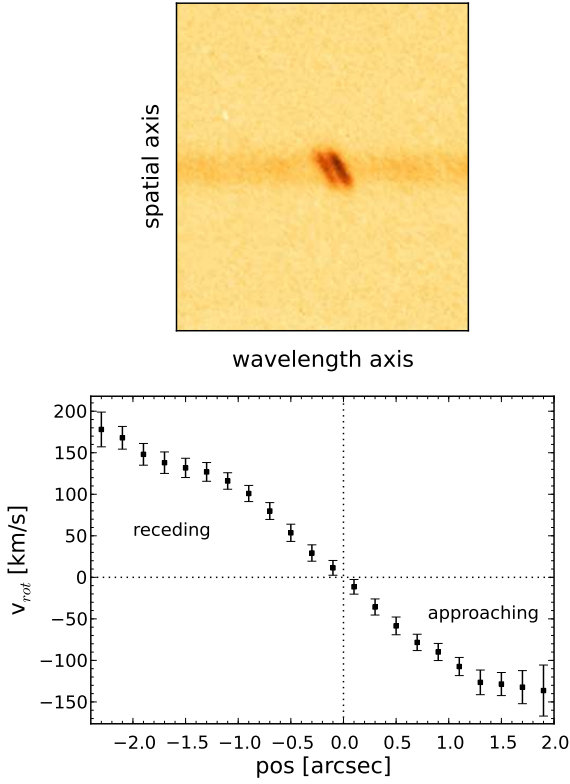}
      \caption{\textit{Upper panel:} The [\ion{O}{II}] doublet of an example galaxy. \textit{Lower panel:} Rotation curve extracted from the above image. The receding and approaching parts of the rotation curve are symmetric. This is an indicator for undistorted kinematics.   
              }
         \label{fig:rc1}
   \end{figure}

First we inspect the one-dimensional spectrum for usable emission lines and define a quality parameter ranging from 0 (= no detectable emission line) to 3 (strong emission line). Then we compute the average in wavelength space of the emission line centred 2D-spectrum to get its profile along the spatial axis. Next we fit a Gaussian to that profile and derive the luminous centre to within 0.1 arcsec . In cases of galaxies with peculiar light profiles, we redefine the centre manually. The luminous centre thus obtained is used to define the kinematic center of a given galaxy.\\
Prior to the emission line fitting we average three neighbouring rows to enhance the S/N. In the case of very weak emission lines, we set this boxcar filter to five rows, corresponding to one arcsecond (spatial scale $=0\farcs205/pixel$). \\
We fit a single Gaussian profile to the emission lines $[\ion{O}{III}] \,\lambda 4959\mathring{A}$, $[\ion{O}{III}] \,\lambda 5007\mathring{A}$ and $H\beta\,\lambda4861\mathring{A}$ and two Gaussians to the $[\ion{O}{II}] \,\lambda 3726/3729$ doublet. For about half of the spectra the single profiles to the [\ion{O}{II}] doublet yield smaller errors and a larger radial extraction range for the rotation curve. Subsequently, we apply automatic row-by-row fits to the emission line at each position along a galaxy's major axis. The red- and blueshifts (with respect to the kinematic centre) along the spectral axis can be transformed into observed line-of-sight rotation velocities and define an observed rotation curve. For more details see \citet{boehm04}.\\ These rotation curves are not yet corrected for the disk inclination $i$. However, in this analysis we are only interested in the quantitative rotation-curve shape, not the normalisation of the rotation velocity.\\
We were able to extract rotation curves from at least one emission line in 182 different galaxies. Of these, 86 are field galaxies and 96 are cluster galaxies. We show an example rotation curve with the corresponding 2D emission line spectrum in Figure \ref{fig:rc1}.

\subsection{Rotation-curve asymmetry}
\begin{figure*}
   \centering
   \includegraphics[angle=0,width=\textwidth]{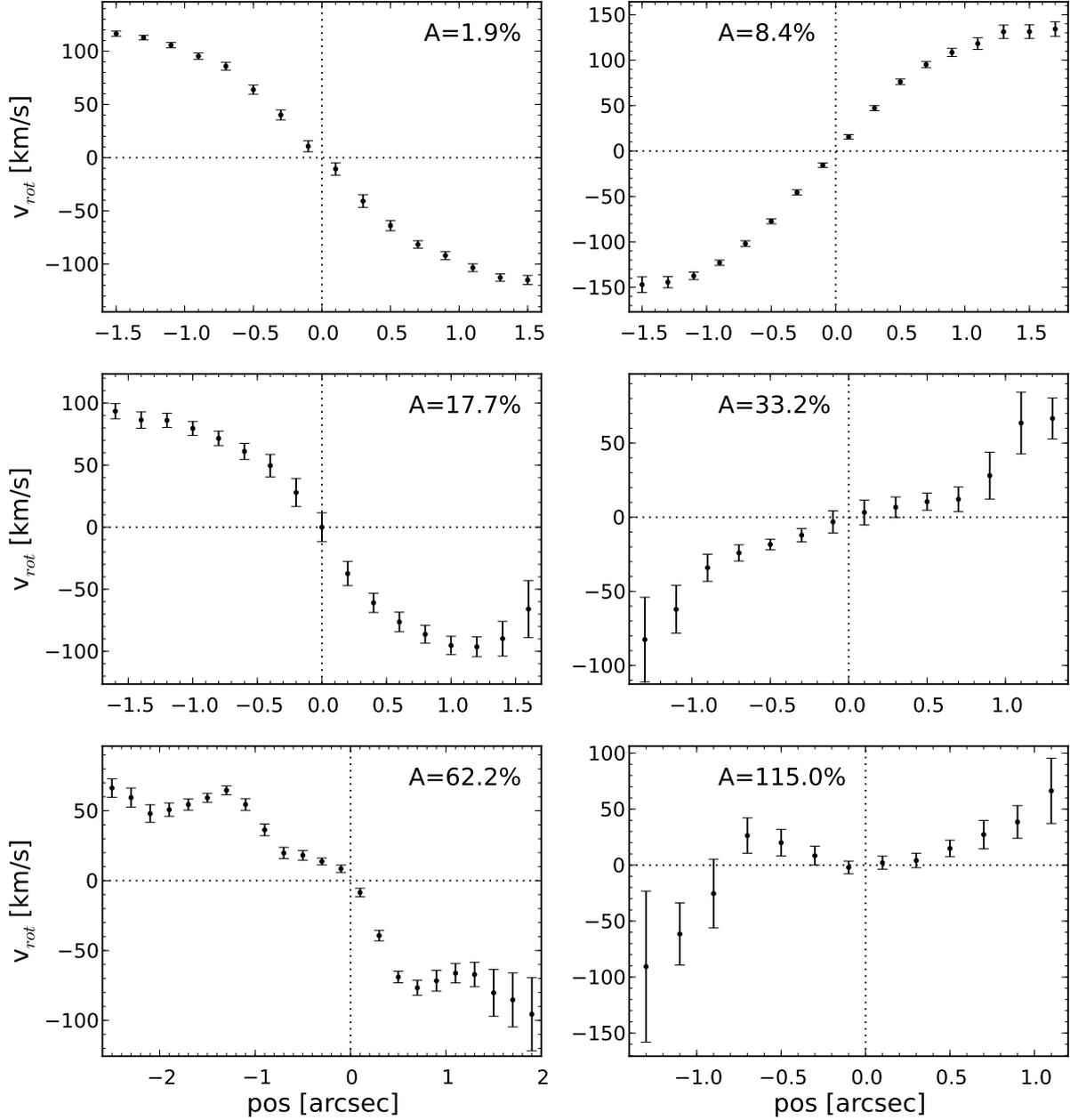}
		\caption{Rotation curves with different degrees of kinematic asymmetry. From top/left to bottom/right the asymmetry index $A$ is increasing.
              }        
\label{fig:test}    
\end{figure*}

\begin{figure}[h]
   \centering
   \includegraphics[angle=0,width=\columnwidth]{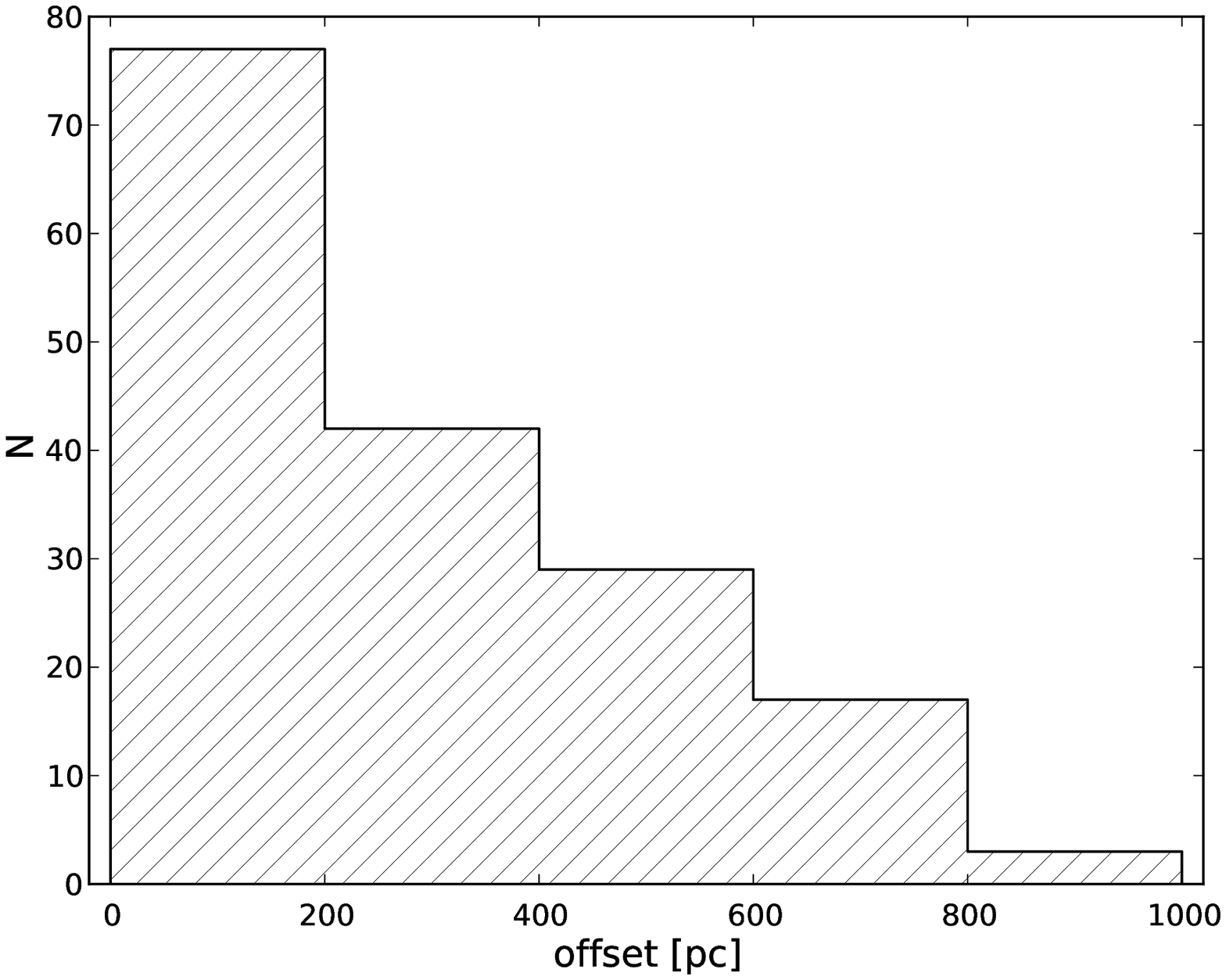}
      \caption{Histogram of the offset between the luminous and kinematic centres, determined by minimising the rotation asymmetry.      
              }
         \label{fig:dckl}
 \end{figure}

\begin{figure}[h]
   \centering
   \includegraphics[angle=0,width=\columnwidth]{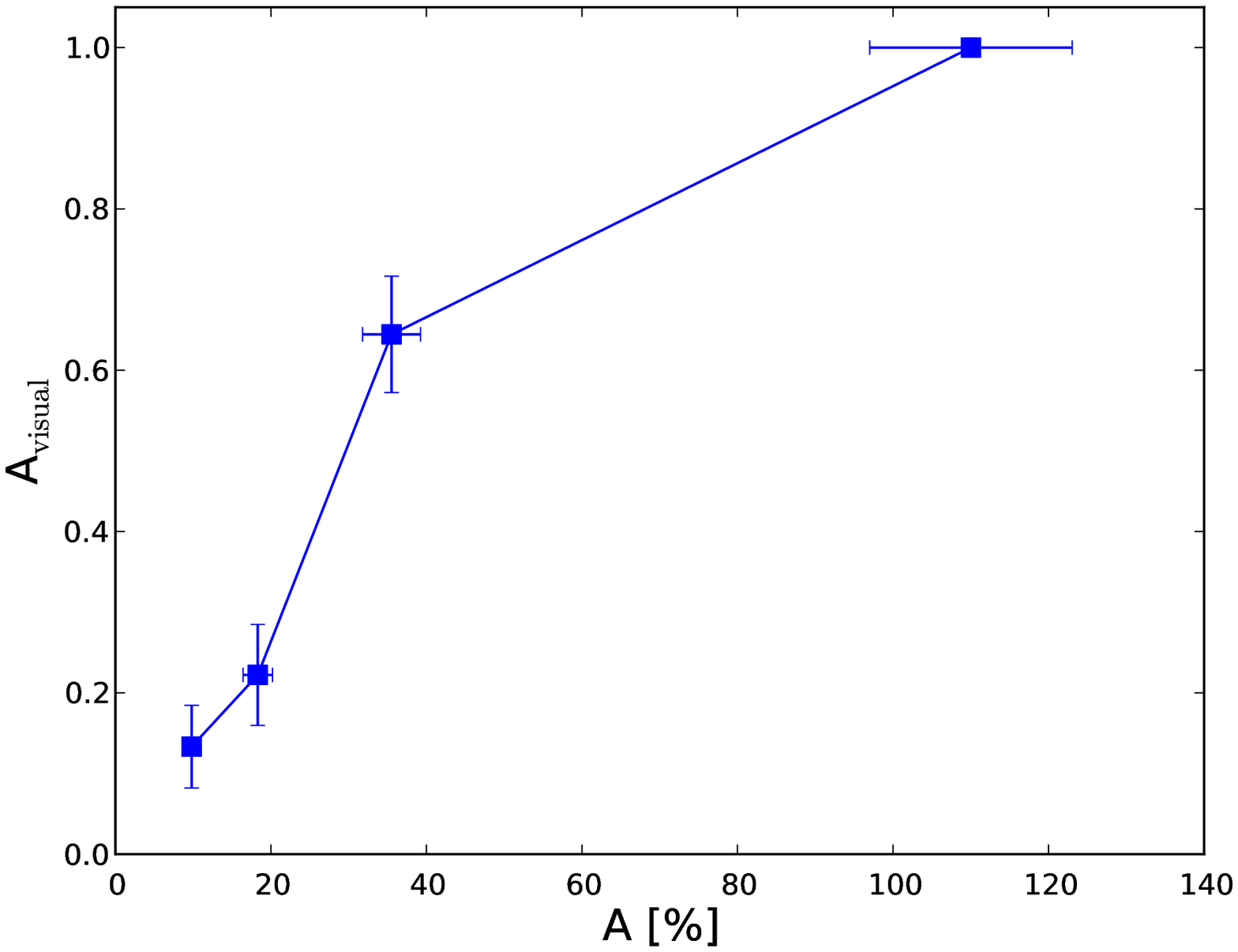}
      \caption{ The visual rotational asymmetry parameter $A_{\mathrm{visual}}$ is plotted vs. the computed asymmetry index $A$. $A_{\mathrm{visual}}$ is 0 or 1 according to whether a rotation curve is considered as undistorted or distorted. The 182 galaxies are combined into bins and than an average is calculated (bin size = 45).
              }
         \label{fig:qv1vsai}
 \end{figure}

To quantify the kinematic distortions of a galaxy we follow the approach of \citet{dale01} and use a similar measure for the rotation-curve asymmetry. We choose
\begin{equation}
      A = \sum_{i}\frac{|v(r_{i})+v(-r_{i})|}{\sqrt{\sigma_{v}^{2}(r_{i})+\sigma_{v}^{2}(-r_{i})}}\cdot \left[ \frac{1}{2}\sum_{i}\frac{|v(r_{i})|+|v(-r_{i})|}{\sqrt{\sigma_{v}^{2}(r_{i})+\sigma_{v}^{2}(-r_{i})}}\right]^{-1}
      \label{eq:asym}
   \end{equation}
The total area between the kinematically folded, approaching and receding side is normalised to the average area under the rotation curve. Additionally, the contribution of each velocity pair ($v(r_{i}),v(-r_{i})$) is weighted by its error ($\sigma_{v}(r_{i}),\sigma_{v}(-r_{i})$). The error $\sigma_A$ of the asymmetry is calculated via error propagation of the velocity errors $\sigma_{v}(r_{i})$. \\
This measure is primarily sensitive to the outer parts of a rotation curve, where the faster velocities imply larger absolute differences in a velocity pair, while the errors remain the same. Accordingly, our asymmetry measure is very sensitive to offsets between the kinematic and luminous centre.\\
The kinematic centre of a galaxy is the position which by definition has no rotational Doppler shift. Since changes in the kinematic centre lead to different asymmetry values, we minimise $A$ by allowing shifts of the assumed kinematic centre by up to $\pm1.5$ pixels with respect to the luminous centre. The centre position that results in the smallest asymmetry value is then set to be the final kinematic centre for all remaining analysis. Also, we inspect the results of the minimisation algorithm visually and treat galaxies with peculiar rotation curves or with a small radial extent individually. In some cases we reduce the allowed shift manually or even set it to 0. \\
For undistorted galaxies this minimising algorithm results in symmetric outer parts of the rotation curve. Figure \ref{fig:test} shows examples of rotation curves with different asymmetry values. Seeing correlates the error bars of adjacent points and it can smooth out rotation-curve asymmetries, especially in smaller galaxies. However, since we do only internal comparisons and focus mostly on asymmetries at large galacto-centric radii, the impact of these effects on our analysis should be small.\\
In Figure \ref{fig:dckl} we show a histogram of the offsets between the kinematic and the luminous centres. \citet{mendes03} adopt the same technique to determine the offset between the kinematic and luminous centre for 25 Hickson compact group galaxies.
The average distance between the kinematic and luminous centre is higher in their compact group galaxies (median = 521pc) than in our cluster galaxies (median = 313pc).  
\\
As a further test, we determine a visual asymmetry parameter $A_{\mathrm{visual}}$ for each visible emission line, and set it to 0 or 1 according to whether a rotation curve is visually considered undistorted or distorted. Figure \ref{fig:qv1vsai} shows the correlation between the computed asymmetry index $A$ and the visual index $A_{\mathrm{visual}}$, which broadly agree.\\

   If for a given galaxy, rotation curves and subsequently asymmetry values $A$ can be determined from more than one emission line, we calculate the final asymmetry value via an error-weighted average of the values of each emission line. 
   
\subsection{Proxies for environment}
One way to define a proxy for the cluster environment is the separation variable $s$ introduced by \citet{carlberg97}, which uses normalised coordinates and is based on the projected position and velocity of a galaxy relative to the cluster centre. Here, we scale velocity differences with respect to each subcluster centre to $\sigma_{\mathrm{scl}}$ and scale projected radial coordinates to $r_{200_{\mathrm{scl}}}$, the radius where the mean interior density is $200\rho_{c}(z)$ (see Table \ref{table:mvir2}): 
 \begin{equation}
      r_s = \frac{r}{r_{200_{\mathrm{scl}}}},
      \label{eq:rs}
   \end{equation}
\begin{equation}
      v_{s} =\frac{v_{\mathrm{rest}}}{\sigma_{\mathrm{scl}}},
      \label{eq:vs}
\end{equation}
where $r$ is the projected radius from the subcluster centre and $v_{\mathrm{rest}}$ is the peculiar rest-frame velocity of a galaxy residing in a certain subcluster. Such a scaling allows to compare galaxies from different subclusters. The separation $s$ is then given by: 
\begin{equation}
      s^{2} = r_{s}^{2}+v_{s}^2
      \label{eq:s}
   \end{equation}
For analysing asymmetry relations we combine objects in bins and compute for each bin an error-weighted average of the individual asymmetry values. This error-weighting is necessary since lower quality rotation curves with large error bars would otherwise bias the asymmetry indices to higher values. The errors of the mean values are calculated taking the maximum of the ``intrinsic error" (error from the $A$-values themselves) and the ``extrinsic error" (error from the scatter of the $A$-values in the bin). The former is usually considered to be the error in the mean and is, for most bins, smaller than the error derived from the scatter around the mean. So, adopting the latter for the error bars is a conservative approach and commonly referred to as a ``correction for overdispersion".

\subsection{Test of selection and segregation effects}
\begin{figure}[]
   \centering
   \includegraphics[angle=0,width=\columnwidth]{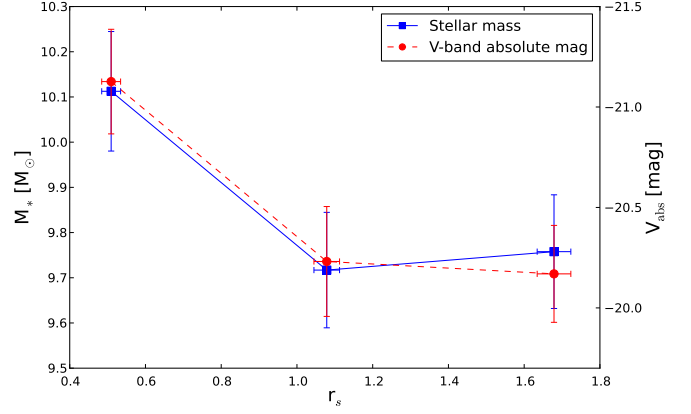}
      \caption{Stellar mass $M_{\ast}$ (blue squares, solid line) and the absolute V-band magnitude $V_{\mathrm{abs}}$ (red circles, dashed line) as a function of the cluster-centric radius $r_s$ for our sample of 96 cluster galaxies (bin size = 32).
              }
         \label{fig:segregation}
 \end{figure}

\begin{figure}[]
   \centering
   \includegraphics[angle=0,width=\columnwidth]{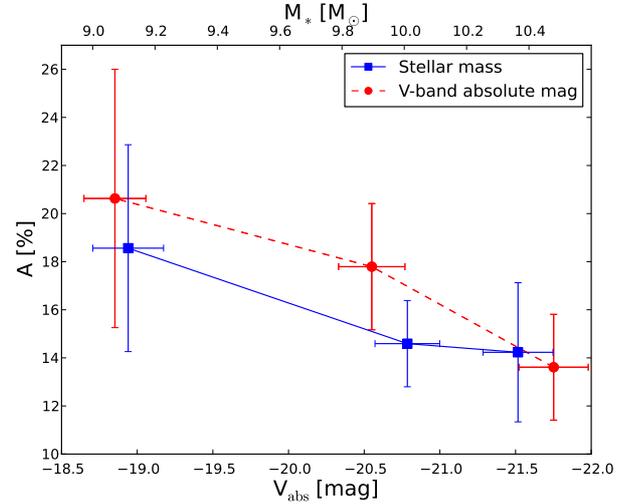}
      \caption{The asymmetry index $A$ as a function of stellar mass $M_{\ast}$ (blue squares, solid line) and absolute V-band magnitude $V_{\mathrm{abs}}$ (red circles, dashed line) for our sample of 96 cluster galaxies (bin size = 32).
              }
         \label{fig:AvsmassV}
 \end{figure}

Before investigating any influence of environment on the rotation-curve asymmetry, we try to consider potential consequences of selection and segregation effects.\\
The well-examined morphology-density relation is expected to be present in our sample. Hence we should detect a mass and luminosity segregation, such that more massive and luminous galaxies reside preferentially near the cluster centres. We rely on absolute magnitudes and stellar masses from the STAGES catalogue \citep{stagescat09} (see Figure \ref{fig:segregation}), and find that  stellar mass and V-band luminosity increase towards smaller cluster-centric radii. Below a radius $r_s\lesssim 0.4$ no spiral galaxies are contained in our sample. This is no surprise since we selected only spiral galaxies with a star-forming SED, which are rare in the cluster cores.\\
Figure \ref{fig:AvsmassV} shows the asymmetry index $A$ as a function of stellar mass $M_{\ast}$ and absolute V-band magnitude $V_{\mathrm{abs}}$ for our sample of 96 cluster galaxies. There is a tendency for low-mass galaxies to have slightly higher rotation-curve asymmetries. The Spearman rank-correlation test, however, shows no significance ($\rho=-0.06$, $p=41.5\%$). Low-luminosity galaxies seem to be slightly more distorted than galaxies with a high V-band magnitude, but also here we find no significant Spearman-test results ($\rho=0.10$, $p=36.2\%$). Since there are anti-correlations they might slightly reduce the average $A$ at small cluster-centric radii. However, it is unlikely that these weak trends have a biasing effect on our results.
 
Our original source catalogue comprises $\sim 320$ cluster galaxies within the selection constraints. Slits could be placed on $103$ cluster members. We could successfully extract rotation curves of $96$ of these cluster galaxies (success rate $\sim 93\%$, i.e. only 7 cluster galaxies selected for spectroscopy yield no rotation curve). Our sub-sample might not be representative of the whole cluster population and differ significantly in some of the properties. There could also be regions in the cluster were the spectroscopic completeness is higher or lower. 
 \begin{figure}[]
   \centering
   \includegraphics[angle=0,width=\columnwidth]{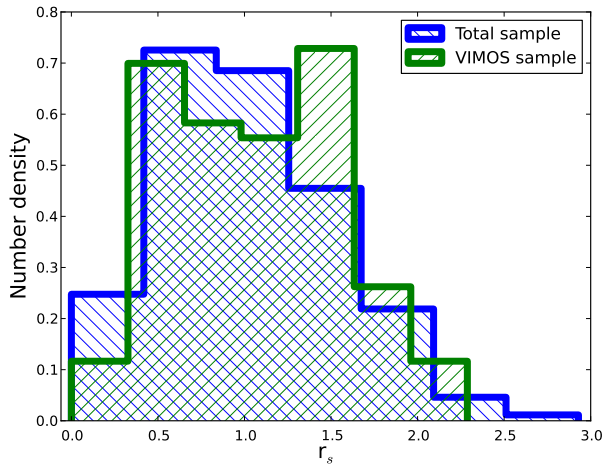}
      \caption{Normalised histograms of the cluster-centric radius $r_s$ for the total target sample (blue, 324 galaxies) and our spectroscopically observed VIMOS sample (green, 103 galaxies).
              }
         \label{fig:ndensity}
 \end{figure} 
Figure \ref{fig:ndensity} shows the normalised histograms of the cluster-centric distance $r_s$ for the total source catalogue and for our spectroscopic sub-sample. The Kolmogorov-Smirnov statistic, which tests whether two samples are drawn from the same distribution, gives a p-value of $47\%$, i.e. we cannot reject the null hypothesis that the distributions are the same. Furthermore, the Mann-Whitney-U test, which  assesses whether one of two groups tends to have larger values than the other, is not significant (p-value = $29\%$). We get similar results if we divide our sample into two luminosity bins or stellar mass bins.\\
Additionally, we investigate the luminosity, stellar mass, star-formation rate and Sersi\'{c} index distribution of each sample. In every case the Kolmogorov-Smirnov test as well as the Mann-Whitney-U test yield non significant differences between source catalogue and spectroscopic sample. Also the variation of aforementioned parameters with scaled cluster-centric distance $r_s$ is the same within the error bars.\\
In summary, these tests clearly indicate that mass (or luminosity) segregation, sample selection effects, spectroscopic incompleteness and other potential biases are unlikely to affect our results significantly.

\subsection{Influence of environment}
\begin{figure*}[]
   \centering
   \includegraphics[angle=0,width=\textwidth]{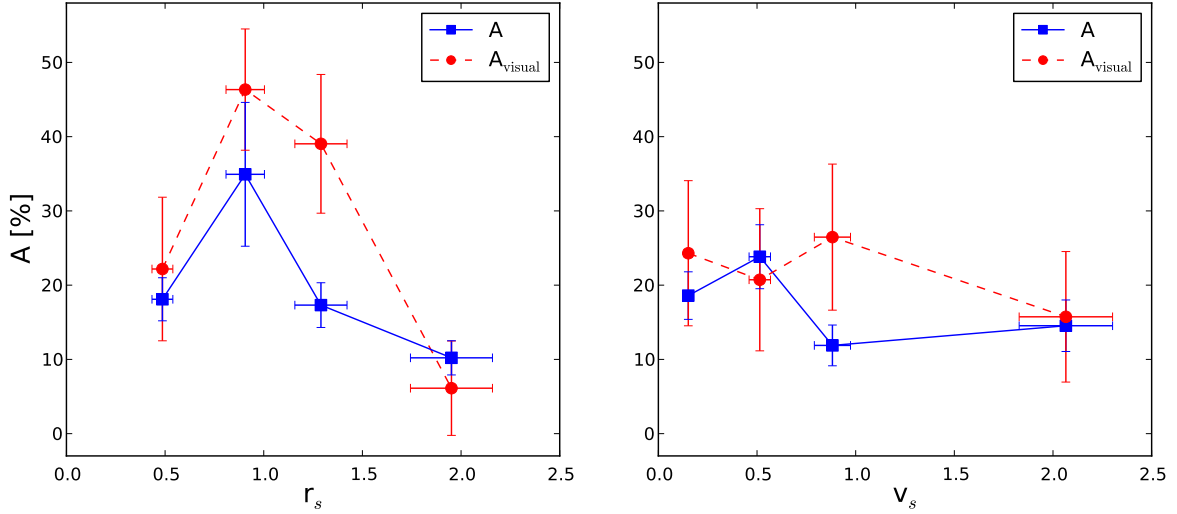}
      \caption{The two asymmetry indices $A$ (blue squares, solid line) and $A_{\mathrm{visual}}$ (red circles, dashed line) versus cluster centric radius $r_s$ (left panel) and peculiar rest-frame velocity $v_s$ (right panel). The asymmetry is significantly higher at intermediate radii and tends to be slightly higher at lower velocities. (96 cluster galaxies; binsize = 24).
           }
         \label{fig:rsvs}
   \end{figure*}

\begin{figure*}[]
   \centering
   \includegraphics[angle=0,width=\textwidth]{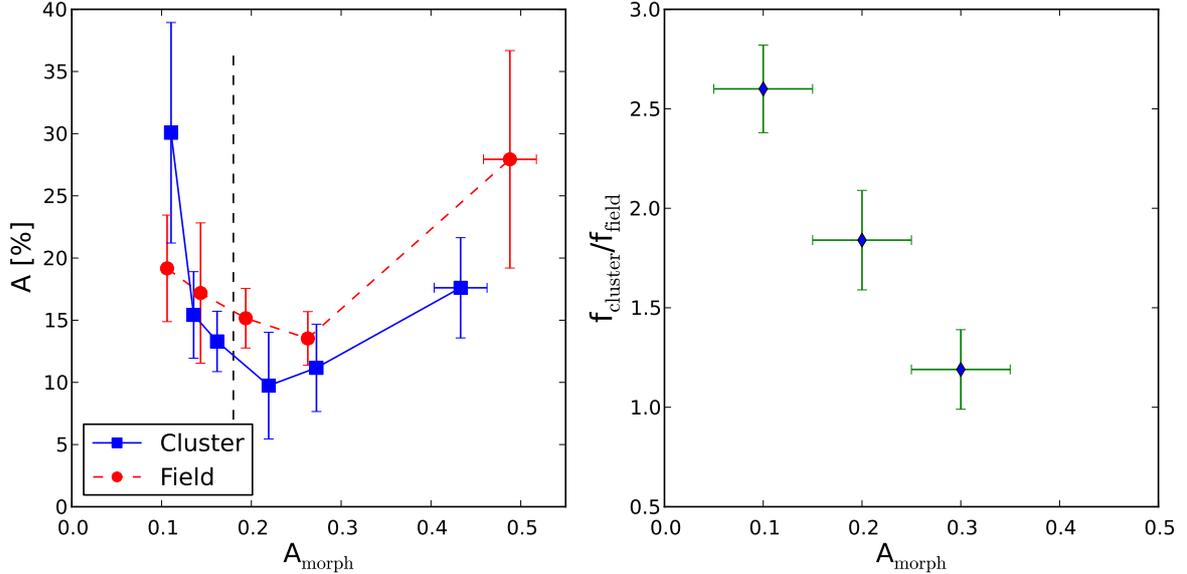}
      \caption{The left panel shows the rotation-curve asymmetry $A$ vs. the morphological asymmetry $A_{\mathrm{morph}}$. The black dashed vertical line at the median $A_{\mathrm{morph}}=0.18$ divides our sample of 96 cluster galaxies into a morphologically distorted group (MDG) ($A_{morph}>0.18$) and a morphologically undistorted group (MUG) ($A_{\mathrm{morph}}<0.18$) (binsize=16). As a comparison we also show the 86 field galaxies (binsize=18). 
      The right panel shows the ratio between the fraction of kinematically distorted cluster $f_{\mathrm{cluster}}$ and field $f_{\mathrm{field}}$ galaxies for three different morphological asymmetry bins.
              }
         \label{fig:avsiacf}
\end{figure*}

\begin{figure*}
   \centering
   \includegraphics[angle=0,width=\textwidth]{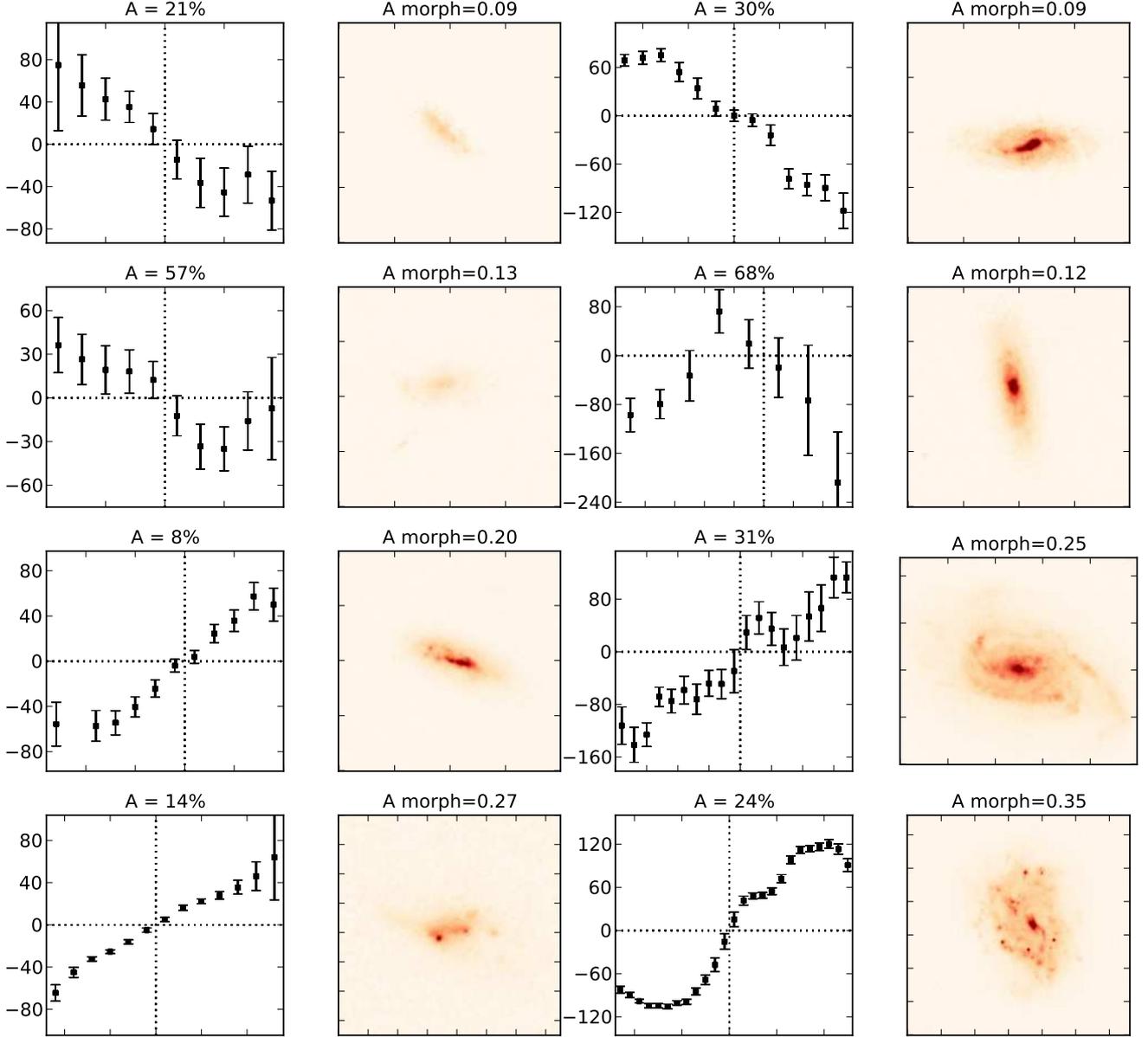}
		\caption{Galaxies with different degrees of morphological asymmetry, $A_{\mathrm{morph}}$, and their corresponding rotation curves.}        
\label{fig:rcimage}    
\end{figure*}

\begin{figure*}[]
   \centering
   \includegraphics[angle=0,width=\textwidth]{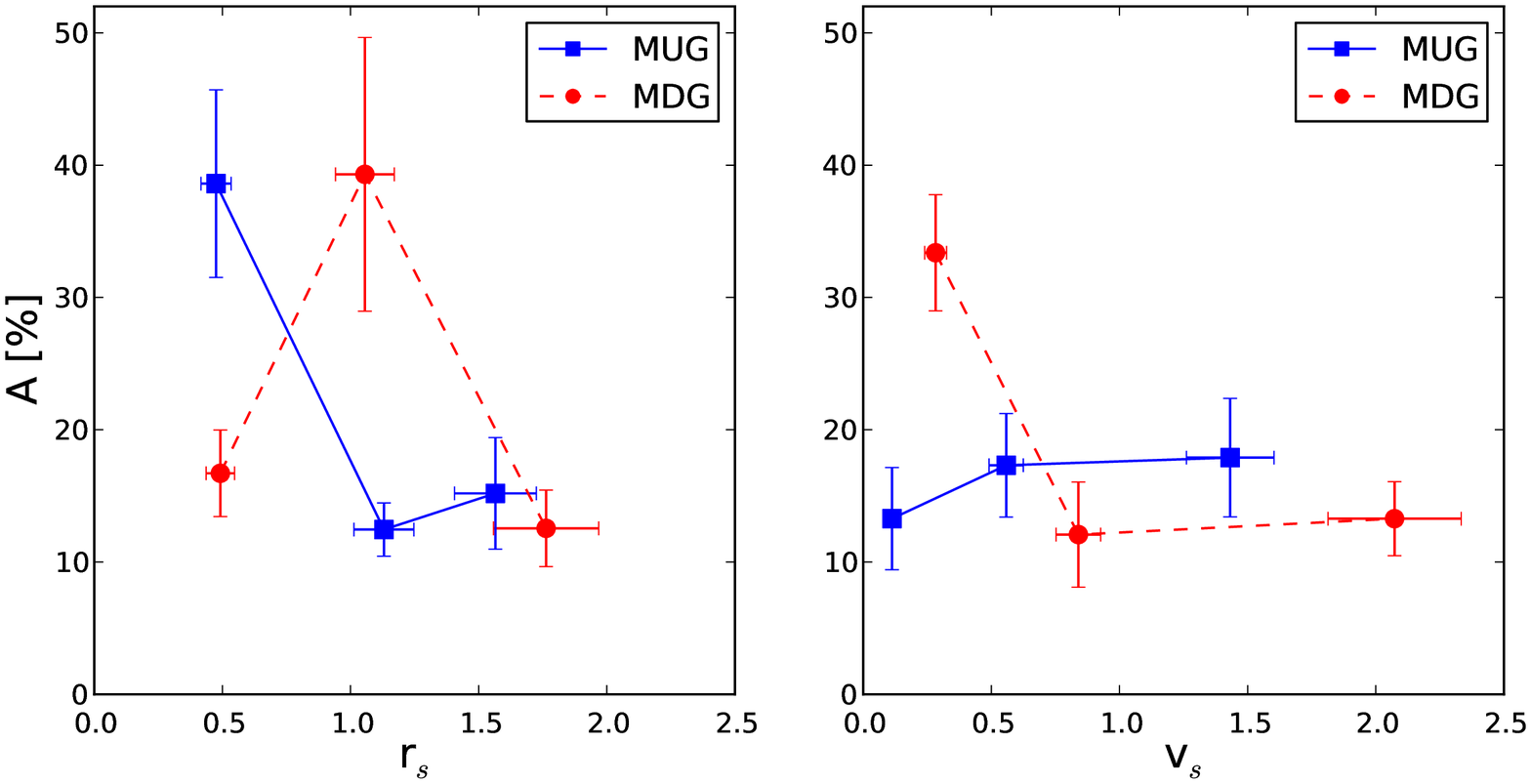}
      \caption{The rotation-curve asymmetry $A$ vs. scaled cluster-centric radius $r_s$ (left panel) and scaled rest-frame velocity $v_s$ (right panel) for MDGs ($A_{\mathrm{morph}}>0.18$, red circles and dashed line) and MUGs ($A_{\mathrm{morph}}<0.18$, blue squares and solid line). Left panel: MDGs have the highest asymmetries at intermediate radii, where tidal interactions are more likely, while MUGs have higher asymmetries near the centre. Right panel: the asymmetry of MDGs peaks at low rest-frame velocities, while MUGs have higher asymmetries for higher values of $v_s$ (binsize=16).
              }
         \label{fig:mgrsvs}
\end{figure*}

As outlined in the introduction, cluster environment plays an important role in the evolution of galaxies. If cluster environment affects the kinematics of galaxies in general or the gaseous disk of spirals in particular it should be related somehow to the asymmetry of a rotation curve.\\
Ram pressure is the force exerted on a galaxy when it moves through the hot ($\sim 10^{8}$ K) and dense ($\sim 10^{-3}$ atoms/$\mathrm{cm^{3}}$) intra-cluster medium \citep{gunn72}. It is stronger at higher relative velocities and at smaller cluster-centric radii, where the ICM density is higher \citep[proportional to $\rho_{\mathrm{ICM}}*v_{\mathrm{rest}}^2$, e.g.][]{gunn72}. A wide range of numerical simulations and observations demonstrate that ram pressure in clusters is sufficient to strip part of the gas from galaxies. The stellar disk, on the other hand, is not disturbed by the ICM, leaving galaxies with a truncated atomic gas (\ion{H}{I}) disk \citep{chung08,kronberger08,boselli06,quilis00,kenney04}.\\
Interactions between a galaxy and the cluster that affect the gaseous halo (aka ``starvation" or ``strangulation") can prevent the replenishment of the gas disk. After consumption of the latter, star formation is quenched \citep[e.g.][]{larson80,balogh00}. In general, such interactions are expected to be more efficient in the dense cluster core.\\
There are also gravitational galaxy-galaxy interactions, that disturb the stellar disk of a galaxy significantly. Frequent high-speed encounters (harassment) heat the stellar component which leads to a more compact morphology due to a decreased rotational velocity and a thicker bulge due to an increased velocity dispersion \citep{moore96}. In cluster cores mergers are rare due to the high relative speeds of galaxies. Instead, they are more likely in the less dense cluster outskirts \citep{boselli06}, where motions are slower.\\
In Figure \ref{fig:rsvs} we show the two asymmetry indices $A$ and $A_{\mathrm{visual}}$ as a function of the cluster centric radius $r_s$ (left panel). We split the 96 cluster galaxies in 4 bins with 24 galaxies each and compute the average asymmetry values in each bin (error-weighted in the case of $A$). The $A_{\mathrm{visual}}$ values are rescaled using the relation in Figure \ref{fig:qv1vsai}. Both asymmetry indices peak at intermediate cluster-centric radii. The Spearman test shows a significant anticorrelation ($\rho=-0.42$, $p=1.7\%$).\\
In the right panel we plot the asymmetries versus the peculiar rest-frame velocity $v_s$. The asymmetry tends to be higher for low peculiar rest-frame velocities, although the correlation is not significant ($\rho=-0.22$, $p=18.8\%$), which indicate a mixture of different processes.\\
\citet{heiderman09} find that tidal interactions and mergers in the A901/902 cluster system dominate in intermediate density regions between the cluster core and the cluster outskirts. The lack of mergers in the innermost regions is most likely due to the large velocity dispersion. Tidal interactions preferably occur at low $v_s$ values, where the encounters between galaxies last longer and the tidal effects can accumulate. In contrast, ram-pressure stripping is stronger at higher rest-frame velocities and at smaller cluster-centric radii, where the ICM-density is higher.\\
\citet{vogt04} investigate optical and \ion{H}{I} properties of cluster and field spirals, and distinguish four different groups of galaxies. Most field galaxies form a \textit{normal} population with no peculiar properties, however, they also find spirals with different kinds of peculiarities, which could represent different evolutionary stages in the transformation of field galaxies into cluster S0 galaxies. \textit{Asymmetric} spirals have distortions mainly on the leading edge of the gas disk. They are probably falling inwards and are on their first transit through the cluster. They are assumed to have less vigorous interactions with the cluster like e.g. ram-pressure stripping. Further classes of \textit{stripped} ($\mathrm{H}\,\mathrm{I}$ deficient; SF confined to 3 disk scale lengths) and \textit{quenched} ($\mathrm{H}_{\mathrm{\alpha}}$ in absorption) galaxies might have already passed through the core and could be transitional stages in the transformation process into an S0 morphology. \citet{vogt04} conclude that gas stripping is a significant process in the evolution of spiral galaxies falling into a cluster and at least contributes to the morphological transformation of spiral galaxies into cluster S0 galaxies.\\
Since a combination of distortions in the rotation curve with a lack of asymmetries in the stellar disk could be an indicator for ram-pressure stripping \citep{kronberger08} we exploit V-band HST-images of our sample galaxies to quantify their morphological asymmetry. \\
We here adopt an asymmetry measure $A_{\mathrm{morph}}$, which is frequently used in the literature \citep[e.g.][]{conselice00}. It is calculated by subtracting a galaxy image $I_{180^{\circ}}$, rotated by $180^{\circ}$, from the original image $I$:
\begin{equation}
      A_{\mathrm{morph}} =\min\left( \frac{\sum_{i}|I-I_{180^{\circ}}|}{\sum_{i}|I|}\right)- \min\left( \frac{\sum_{i}|B-B_{180^{\circ}}|}{\sum_{i}|I|}\right)
      \label{eq:ia}
\end{equation}
$B$ and $B_{180^{\circ}}$ are defined analogously to account for the contribution from sky noise. $B$ and $B_{180^{\circ}}$ are blank sky regions of the same size as the original galaxy image. The sums are calculated over all pixels within the $1\sigma$-isophote of a galaxy. To avoid an influence from the determination of the centre of a galaxy, we compute a minimisation by shifting the galaxy image on a pixel grid, resulting in typical shifts of 2-3 pixels ($\stackrel{\scriptscriptstyle\wedge}{=} 0\farcs 06-0\farcs 09$) \citep{boehm12a}.\\
Figure \ref{fig:avsiacf} (left panel) shows the rotation-curve asymmetry $A$ vs. the morphological asymmetry $A_{\mathrm{morph}}$ for cluster and field galaxies. For several interaction processes (like galaxy-galaxy interactions, harassment) we expect coinciding morphological and kinematic distortions. However, we find just a slight increase in the rotation-curve asymmetry $A$ towards higher morphological asymmetry $A_{\mathrm{morph}}$. On the other hand there seems to be a significant fraction of morphologically undistorted but kinematically distorted cluster galaxies.
The threshold $A_{\mathrm{morph}}=0.18$ (vertical line in Figure \ref{fig:avsiacf}) is the median of our sample and divides the 96 cluster galaxies into  morphologically distorted (MDGs) ($A_{\mathrm{morph}}>0.18$) and  morphologically undistorted galaxies (MUGs) ($A_{\mathrm{morph}}<0.18$). Applying a Spearman rank-order correlation test to the class of MUGs, we find a significant anticorrelation ($\rho=-0.41$, $p=6.8\%$). On the other hand we get no significant result for the class of MDGs ($\rho=0.05$, $p=43.8\%$).\\
As a test we apply the same analysis to the 86 field galaxies. In contrast to the cluster sample, the peak of $A$ for low $A_{\mathrm{morph}}$ is missing. Moreover, the Spearman test above and below the median gives no significant result for the field galaxies (MUG: $\rho=0.05$, $p=43.6\%$; MDG: $\rho=0.24$, $p=26.2\%$). This indicates that the peak at high $A$ and low $A_{\mathrm{morph}}$ for cluster galaxies results from a cluster-specific interaction process. MDGs might be dominated by tidal interactions while morphologically undistorted cluster galaxies with perturbed rotation curves, might be subject to ram-pressure stripping, an effect that is not expected in the field environment.\\ 
\begin{figure*}[]
   \centering
   \includegraphics[angle=0,width=\textwidth]{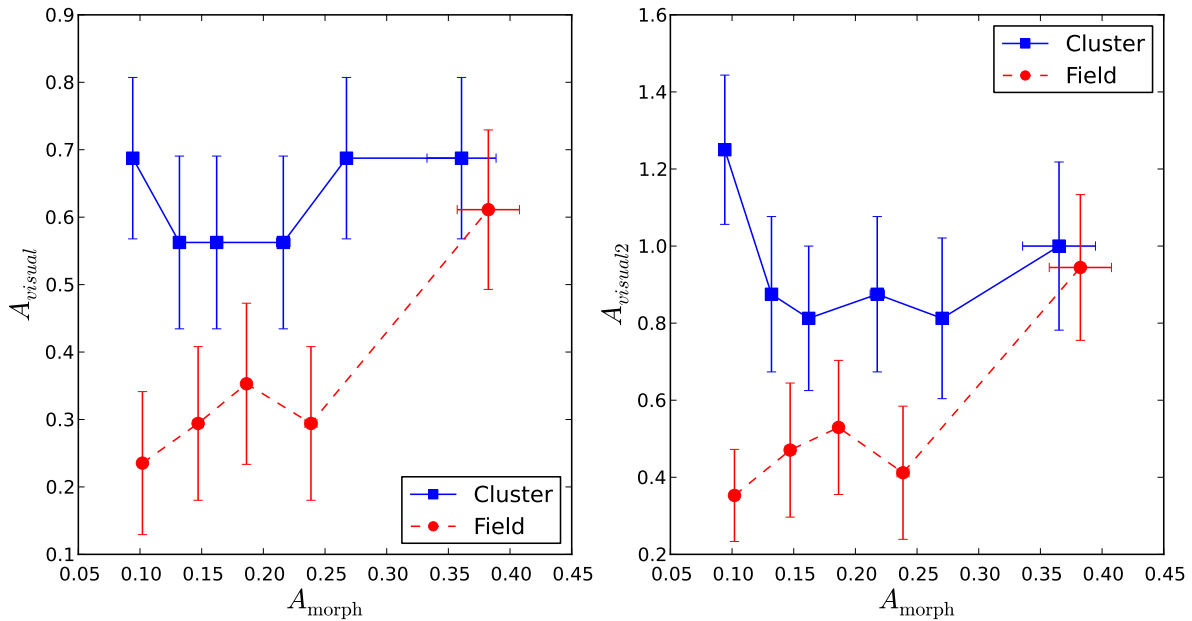}
      \caption{The visual rotation-curve asymmetry vs. the morphological asymmetry $A_{\mathrm{morph}}$ for cluster (binsize=16) and field (binsize=17) galaxies. $A_{\mathrm{visual}}$ (left panel) classifies galaxy kinematics as undistorted (0) or distorted (1), while  $A_{\mathrm{visual 2}}$ (right panel) differentiates between three classes: undistorted (0), slightly distorted (1) and heavily distorted (2).
              }
         \label{fig:avsiacfav}
\end{figure*} 
Figure \ref{fig:avsiacfav} (left panel) is similar to Figure \ref{fig:avsiacf} but uses the visual rotation-curve asymmetry $A_{\mathrm{visual}}$. Here, in the cluster the fraction of distorted galaxies ($63\pm5\%$) is significantly higher than in the field sample ($36\pm5\%$). Moreover, morphologically distorted field galaxies are more likely to also show disturbed kinematics ($\rho=0.25$, $p=1.3\%$). In other words, distortions of the gaseous disk in the field are preferentially accompanied by disturbances of the stellar disk. This might be due to the low relative velocities of galaxies in the field, making tidal interactions more efficient. Interestingly, the anticorrelation of MUGs in the cluster is no longer significant when considering the visual asymmetry parameter. This might suggest that, for MUGs, it is not the fraction of kinematically distorted galaxies that increases significantly, but rather the intensity of their distortions. To verify this conclusion we introduce a refined classification parameter, $A_{\mathrm{visual 2}}$ (see Figure \ref{fig:avsiacfav}, right panel), that differentiates between three classes of kinematics: undistorted (0), slightly distorted (1) and heavily distorted (2). Using this finer scheme we find again a significant anticorrelation for morphological undistorted cluster galaxies ($\rho=-0.24$, $p=4.5\%$). Figure \ref{fig:avsiacf} (right panel) clarifies this conclusion. Here we show the ratio between the fraction of kinematically distorted cluster galaxies, $f_{\mathrm{cluster}}$, and field galaxies, $f_{\mathrm{field}}$, for three different morphological asymmetry bins. For morphologically undistorted cluster galaxies a mechanism comes into play that increases the kinematic asymmetry.\\
To summarise this part of our analysis, we found evidence for a cluster-specific interaction process that only effects the gas kinematics, not the stellar morphology. Ram-pressure stripping is, therefore, a prime candidate. \\ 
To illustrate different asymmetry classes, Figure \ref{fig:rcimage} shows HST V-band images with increasing degrees of $A_{\mathrm{morph}}$ and their corresponding rotation curves.\\
In Figure \ref{fig:mgrsvs} we show the asymmetry of MUGs and MDGs as a function of $r_s$ and $v_s$, respectively. We find clear evidence for important galaxy populations with distinct characteristics.  MUGs might be affected by ram-pressure stripping, since they exhibit their asymmetry values are anticorrelated with $r_s$ for $r_s\lesssim 1.1$ ($\rho=-0.85$ and $p=0.4\%$) and positively correlated with their rest-frame velocity ($\rho=0.34$, $p=13.0\%$). Both the high ICM density expected at low $r_s$ and high relative velocity will boost ram pressure and, consequently, produce larger kinematic asymmetries.\\
On the other hand, MDGs are most likely dominated by mergers and tidal interactions. The asymmetry index $A$ peaks at intermediate radii ($r_{s}\sim1.0$) and is higher for lower rest-frame velocities. The Spearman test at greater radii confirms a significant anticorrelation ($\rho=-0.45$, $p=3.4\%$). At intermediate radii, the galaxy density is already high and the relative galaxy velocities are lower than in the cluster core, making tidal interactions more efficient. \\

\subsection{Dusty red galaxies}
\begin{figure}[]
   \centering
   \includegraphics[angle=0,width=\columnwidth]{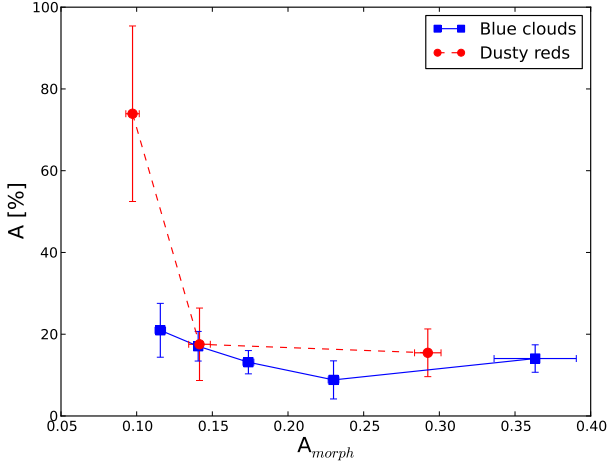}
      \caption{Same as Figure \ref{fig:avsiacf}, but considering 28 dusty red (red circles, dashed line; binsize=9) and 68 blue cloud (blue squares, solid line; binzise=12) cluster galaxies, separately.  
              }
         \label{fig:avsiadrbc}
\end{figure}

\begin{figure}
   \centering
   \includegraphics[angle=0, width=\columnwidth]{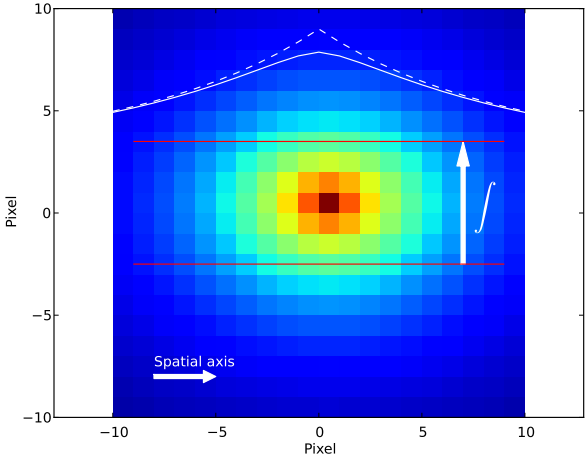}
      \caption{An illustration of the flux integration perpendicular to the slit direction. The exponential profile is projected due to the inclination, $i$, of the disk (here  $i=40^{\circ}$). Hence, the isocontours have elliptical shapes. The galaxy and slit dimensions shown here are typical of our data. In the upper part, the integrated and PSF-convolved profiles (solid line) are compared with an exponential function (dashed line).}
         \label{fig:islit}
\end{figure}

\begin{figure}
   \centering
   \includegraphics[angle=0, width=\columnwidth]{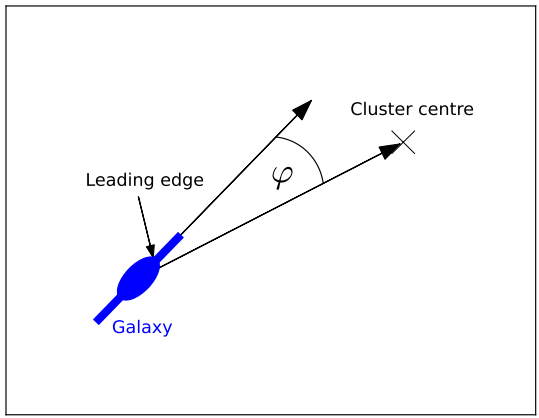}
      \caption{We define the projected infall angle $\varphi$ as the angle between the apparent major axis of the galaxy and the radius vector pointing towards the cluster centre. The galaxy edge closer to the cluster centre is assumed to be the leading edge.}
         \label{fig:phisketch}
\end{figure}

\begin{figure*}
   \centering
   \includegraphics[angle=0, width=\textwidth]{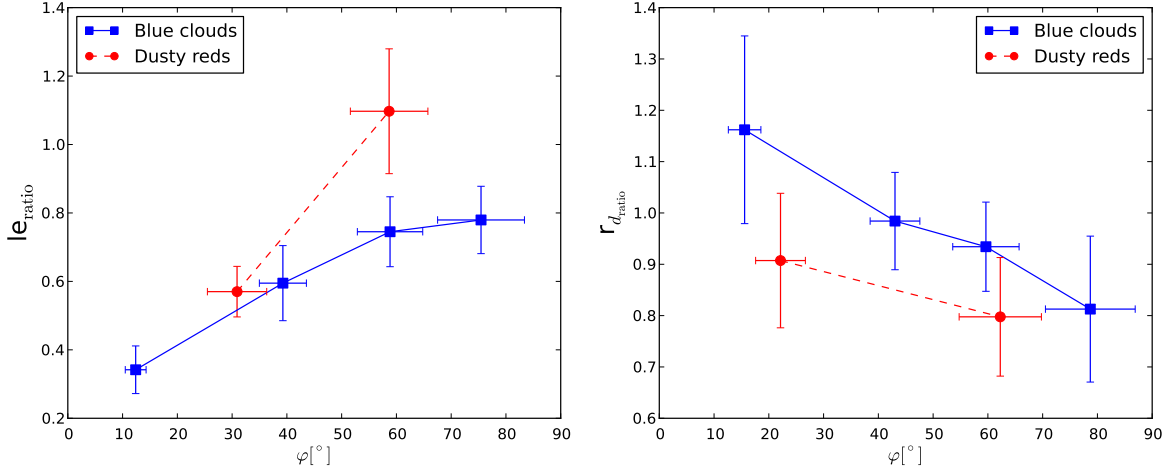}
      \caption{The leading edge ratio $le_{\mathrm{ratio}}$ (left panel) and scale length ratio $r_{d_{\mathrm{ratio}}}$ (right panel), respectively,  are shown as a function of the projected infall-angle  $\varphi$ for blue cloud (blue squares, solid line) and dusty red galaxies (red circles, dashed line). Left panel: the leading edge ratio decreases as the infall becomes more edge-on. Right panel: the scale length ratio declines for a more face-on infall (binsize: blue clouds=15, dusty reds=10).}
         \label{fig:phi}
\end{figure*}

In this section we consider dusty red galaxies (selection criterion in the public catalogue \citet{stagescat09}: $\mathrm{sed\_type} = 2$) more closely.  Twenty-eight of the 96 cluster galaxies in the catalogue are assigned that SED type. Although they are actively star-forming, these galaxies appear red. Their colours are a combined effect of i) an (on average) four times lower SFR than blue cloud galaxies, and ii) intrinsic dust extinction. Having extinction levels similar to blue cloud galaxies, dusty red galaxies can be understood as the low specific-SFR tail of the blue cloud \citep{wolf03}. While a fit to the colour-magnitude relation (CMR) of the red sequence and a parallel cut on its blue side (0.25 mag) are used to differentiate between red and blue galaxies, red galaxies with a higher dust reddening then  $E_{\mathrm{B-V}}=0.1$ are called \textit{dusty red} \citep{wolf05}. They appear predominantly in the medium-density outskirts of clusters, while being rare in higher and lower density regions. \citet{wolf05} suggest two main possible origins.  In one scenario, dusty red galaxies representa a transition stage between blue field and cluster S0 galaxies.  The other possibility is that a dusty red galaxy is the product of a minor merger between an old red cluster galaxy and an infalling blue field galaxy.\\
\citet{jaffe11b} conclude from studies of the intrinsic colour scatter, where they find no significant evolution up to $z\sim0.8$, that most cluster elliptical and S0 galaxies have already joined the red sequence when the ``final" morphology is established. \citet{bamford09} detect a considerable fraction of spiral galaxies with red colours in denser environments, and a substantial population of early-type galaxies with blue colours in low-density environments. This implies a dependence of colour on environment beyond the morphology-density relation. Consequently, the transformation of galaxies from blue to red colours must occur on shorter time-scales than any transformation from spiral to early-type morphology. These findings favour the first scenario proposed by \citet{wolf05}, in which dusty red galaxies are progenitors of S0 galaxies.\\
We now investigate whether there are differences in the properties of dusty red and blue cloud galaxies. The dusty red fraction in MUGs is about $8\%$ higher than in MDG, indicating that the morphological asymmetry is slightly lower. Similar to Figure \ref{fig:avsiacf} we analyse the rotation-curve asymmetry $A$ as a function of the morphological asymmetry $A_{\mathrm{morph}}$, for each SED type. Figure \ref{fig:avsiadrbc} shows that there is a dusty red galaxy population with strong kinematic distortions but no significant morphological disturbances. The Spearman test for morphological undistorted dusty red galaxies ($A_{\mathrm{morph}}<0.18$) confirms a significant anticorrelation ($\rho=-0.68$, $p=2.5\%$), while the correlation for blue cloud galaxies is weaker and less significant ($\rho=-0.40$, $p=17.6\%$). This favours the scenario that dusty red galaxies are more exposed to ram-pressure stripping than blue cloud galaxies. We stress that many dusty red galaxies with high values of $A_{\mathrm{morph}}$ are seen at high disk inclination angles, i.e. they are ``pseudo-asymmetric" due to the prominent dust disk.
To better understand Figure \ref{fig:avsiadrbc} we investigate the spatial distribution of the emission line flux, which originates from the gaseous disk of the galaxy and is therefore much more likely to be affected by ram-pressure stripping than the stellar disk \citep[see e.g.][]{kronberger08}. A measure for the concentration of the line emission is the spectroscopic scale length $r_{d_{\mathrm{spec}}}$, determined via an exponential fit to the emission line flux $I$ along the spatial axis of the 2D spectra:
\begin{equation}
      I(r) = I_{0}\;e^{-\frac{r}{r_d}}.
      \label{eq:rd}
\end{equation}
Prior to fitting the data, we subtracted the continuum emission (estimated red- and blueward of the emission line) to reduce flux contamination from the stellar disk. In the wavelength direction we averaged flux within a $2 \AA$ interval around the emission line. The centre of the galaxy was defined to be located at the flux maximum. In a first step, we performed an exponential fit on each side of this maximum. These values were later used to infer the gas disk asymmetry.\\
However, to derive the total scale length of a galaxy we used a more sophisticated fitting process, in which we considered the effects of the slit width (in the radio bands often referred to as ``beam smearing"), the Point Spread Function (PSF), and the inclination, $i$, of the galaxy with respect to the line of sight. All three effects would cause an overestimation of the scale length when unaccounted for. \\
 To correct for the projection effect due to the inclination of the disk, the 2D exponential profile was modified in the following way:
 \begin{equation}
      I(x,y) = I_{0}\,\exp(\frac{-\sqrt{(x-x_{0})^{2}+(\frac{y-y_{0}}{q})^{2}}}{r_{d}})\;,
      \label{eq:prd}
\end{equation}
 where $(x_0,y_0)$ are the coordinates of the galaxy center and q denotes the ratio of the apparent minor and major axes of the galaxy. The values of these axial ratios were previously determined using the SExtractor package \citep{bertin96} on HST/ACS V-band images, and were adopted from the STAGES master catalogue \citep{stagescat09}.\\
 Since the slits were aligned along the apparent major axes of the galaxies, the projected exponential profile (Equation \ref{eq:prd}) can be integrated perpendicular to the slit direction to get the corresponding 1D-profile. Figure \ref{fig:islit} illustrates how this integration was done. 
The final fitting function was obtained by convolving this 1D-profile with the PSF. The PSF was assumed to be a Gaussian with a FWHM determined from the mean DIMM seeing values during the spectroscopy.\\
Finally, we correct the best-fit scale length values for the position angle $\theta$, which is the tilt angle of the apparent major axis with respect to the spatial axis in the 2D spectrum.\\
Distorted galaxies have larger fitting errors in their $r_{d_{\mathrm{spec}}}$ values. We flag flux emission profiles that are too peculiar, or for which the fit failed, and do not use them in the following analysis.  This excludes 25 out of 96 cluster galaxies, and 11 out of 86 field galaxies.\\
The photometric scale length $r_{d_{\mathrm{phot}}}$ is a measure of the concentration of the continuum emission, which stems from the stellar disk. We derived the values of the photometric scale length using the GALFIT package \citep{peng02} on the corresponding HST/ACS V-band images. In the course of this, we performed a bulge/disk decomposition of the galaxies. We assumed a de Vaucouleurs profile for the bulge and an exponential profile for the disk.\\
We normalised the gas disk scale length to the stellar disk scale length, by computing the scale length ratio $r_{d_{\mathrm{ratio}}}=r_{d_{\mathrm{spec}}}/r_{d_{\mathrm{phot}}}$. When values from more than one emission line were available, we computed an error-weighted average.\\
We find no significant differences in the mean scale length ratios of different SED types. Dusty red galaxies have a mean $r_{d_{\mathrm{ratio}}}$ of $0.86\pm0.08$, while for blue cloud galaxies we find a value of $0.99\pm0.07$. All cluster galaxies combined show a mean scale length ratio of $0.95\pm0.05$, which is about $25\%$ lower than the value for the field galaxies ($1.27\pm0.07$). This lower average scale length ratio of cluster galaxies is most likely the result of quenching in progress (removal of the gaseous halo and faster consumption of the gas disk) or ram pressure. Note that our cluster galaxies are on average $0.27$ dex more massive than our field galaxies, a difference already evident in the source catalogues.  We find no significant correlation between mass and scale length ratio, which implies a mechanism beyond the morphology-density relation.\\
This finding agrees with \citet*{bamford07}, who found that cluster galaxies on average have a $25\%$ lower scale length ratio than field galaxies. Also \citet{phdmj03} found higher scale length ratios for field galaxies than for cluster galaxies. In contrast to this, \citet{jaffe11} found scale length ratios of $\sim 0.8$, independent of environment.\\
Our results suggest that the interactions of the gaseous disk of cluster spirals with the ICM, with other galaxies, or with the cluster potential result in more centrally concentrated star formation. Since the effect is stronger for dusty red galaxies, it is likely that they have been exposed to this environment in a stronger way, or at least for a longer period. A more centrally concentrated star formation might also be important for the growth of the bulge component. Besides the depletion of the gas, this is a further requirement for the transformation of a spiral galaxy into one with an S0 morphology. This scenario is supported by the bulge-to-total ratios obtained from the bulge/disk decompositions using GALFIT. While field galaxies ($0.086\pm0.013$) and cluster blue cloud galaxies ($0.091\pm0.017$) have similarly low bulge-to-total ratios, dusty red galaxies have, on average, a ratio of $\langle B/T \rangle=0.232\pm0.039$.\\
The position angle (orientation of the apparent major axis) of a given galaxy on the sky can be used to introduce a raw estimate of the 3-D angle of the galaxy's movement through the ICM. We define this projected ``infall-angle" $\varphi$ as the angle between a galaxy's apparent major axis and a vector connecting the galaxy and the nearest subcluster center (see Figure \ref{fig:phisketch}). We are aware that, for a given object, this angle $\varphi$ might differ strongly from the real 3-D movement angle. However, in a statistical sense, it should be reliable to use $\varphi$ to get a rough estimate of the angle under which the ICM acts on an infalling galaxy's gas disk. $\varphi$ spans the interval $[0^{\circ},90^{\circ}]$, where $0^{\circ}$ corresponds to edge-on and $90^{\circ}$ corresponds to face-on infall.\\
We also define a ``leading edge ratio" $le_{\mathrm{ratio}}$. This ratio compares the gas scale length (determined from the 2-D spectra) on one side of the disk to that of the other. We compute the $le_{\mathrm{ratio}}$ by dividing the gas scale length on the side of the disk that (in projection) points towards the cluster center (leading edge) by the gas scale length on the side of the disk that (in projection) points away from the cluster center (trailing edge). It will be interesting to see whether the projected infall angle $\varphi$ has an impact on the $le_{\mathrm{ratio}}$, which is a proxy for gas disk asymmetry.
\\
Figure \ref{fig:phi} (left panel) shows the leading edge ratio $le_{\mathrm{ratio}}$ as a function of the projected infall angle $\varphi$ for blue cloud and dusty red galaxies. For more edge-on galaxies (smaller $\varphi$) the leading edge ratio is reduced, implying that ram pressure is successively depleting gas from the leading edge. The Spearman test for all cluster galaxies confirms a highly significant correlation ($\rho=0.49$, $p=0.3\%$). Furthermore, this clear dependence on the infall angle suggests that most galaxies are indeed on their first passage through the cluster. This agrees with the aforementioned class of \textit{asymmetric} spirals \citep{vogt04}. The authors find that these galaxies have distortions mainly on the leading edge of the gas disk, are probably falling into the cluster, and are subject to less vigorous interactions (such as ram pressure stripping) with the cluster.\\
The right panel of Figure \ref{fig:phi} shows the scale length ratio $r_{d_{\mathrm{ratio}}}$ as a function of the projected infall angle $\varphi$. We see a slight reduction of $r_{d_{\mathrm{ratio}}}$ for more face-on galaxies (higher $\varphi$).  However, this trend is not significant ($\rho=-0.14$, $p=13.2\%$). Such a reduction is seen in simulations.  For example, \citet[][]{steinhauser12} found in their combined N-body/hydrodynamic simulations of ram-pressure stripping that the amount of stripped gas is larger when the infall orientation is more face-on.

\begin{figure}
   \centering
   \includegraphics[angle=0, width=\columnwidth]{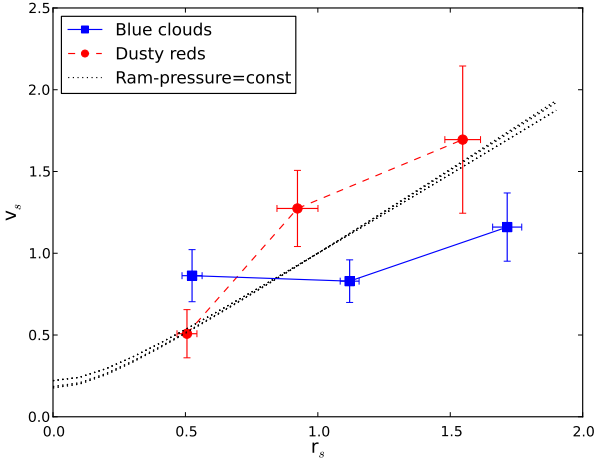}
      \caption{Scaled peculiar rest-frame velocity $v_s$ is shown as a function of the scaled cluster-centric radius $r_s$ for blue cloud (blue squares, solid line) and dusty red galaxies (red circles, dashed line). High-velocity dusty red galaxies are predominantly residing in the cluster outskirts (binsize: blue clouds=22, dusty reds=9). The black dotted lines indicate lines of constant ram pressure for each subcluster, normalised to $v_s$=$r_s$=1.}
         \label{fig:vsvsrsall}
\end{figure}

Figure \ref{fig:vsvsrsall} shows the scaled rest-frame velocity $v_s$ as a function of the scaled cluster-centric radius $r_s$. While blue cloud galaxies show no significant correlation ($\rho=0.16$, $p=11.2\%$), dusty red galaxies have higher velocities at large radii ($\rho=0.46$, $p=0.7\%$). In contrast to our dusty red and blue cloud sample, early type galaxies have, in general, higher velocities at smaller radii, due to the deeper gravitational potential well towards the centre. This discrepancy might be partly due to a high fraction of infalling and ``non-virialised" galaxies. The strongly peculiar relation for our dusty red galaxies could be explained within the framework of ram pressure stripping. If we assume that most of the dusty red galaxies are falling into the cluster (which is supported by Figure \ref{fig:phi}) this finding could best be explained as a selection effect, since ram pressure is proportional to $\rho_{\mathrm{ICM}}*v_{\mathrm{rel}}^2$ \citep[e.g.][]{gunn72}. Under the assumption that dusty red galaxies are undergoing (and/or produced by) strong ram pressure, galaxies residing in the outskirts of the cluster need a high relative velocity to compensate for the low ICM density. On the other hand, for dusty red galaxies with high rest-frame velocities residing close to the cluster centre, ram pressure might be so high that these objects are already quenched, with no detectable star formation. Since we have selected our galaxies by a star-forming SED, these quenched objects with high velocities at low radii are missing in our sample.\\
For the Virgo cluster, \citet{chung08} found stripped H I disks outside the core region ($>1\mathrm{Mpc}$), some of them showing signatures of ongoing ram-pressure stripping and providing evidence that ram pressure is also effective at such distances. Exploiting the dark matter properties of the weak lensing measurements \citep{heymans08}, we follow \citet{makino98} and estimate the distribution of the gas in each subcluster. Assuming an isothermal $\beta$-model for the intra-cluster medium \citep{cavaliere76}, we get ICM-density estimates that are $\rho_{\mathrm{gas}}\sim10^{-28}$ $\mathrm{gcm^{-3}}$ at $1.5*r_{200}$, high enough to produce significant ram pressure if galaxies have relative velocities around $1000$ $\mathrm{kms^{-1}}$ as has been shown with simulations \citep{steinhauser12}. Furthermore, \citet{kenney04} suggested that a merging of subclusters (which is the case for A901/902) can cause large bulk motions and local density boosts, raising the ICM pressure temporarily. This is also seen in simulations by \citet{tonnesen07}, and it facilitates ram-pressure stripping in the outskirts. These ICM density boosts might not show up on shallow X-ray maps like our XMM data of A901/902.\\
At a given relative velocity, the time scale for a galaxy to lose all of its gas due to ram pressure will depend on the ICM density. That is, it is less likely to observe a relatively fast galaxy that still has some gas disk reservoir closer to the cluster core. This could explain the decrease in high velocity dusty red galaxies at smaller cluster-centric radii. For blue cloud galaxies this effect might be less severe, due to higher gas mass fractions when falling into the cluster system. To strengthen this conjecture, the black dotted lines in Figure \ref{fig:vsvsrsall} indicate the regime of constant ram pressure in the $r_{s}$-$v_{s}$ plane. The constant is set to be the ram-pressure value at $r_{s}$=$v_{s}$=$1$, i.e. $\mathrm{const}=\rho_{\mathrm{gas}}(r_s$=$1)*v_{\mathrm{rel}}(v_s$=$1)^{2}$. Since the $\beta$-profile scales with $1/r^{2}$ for radii well outside the core-radius, $v^{2}/r^{2}=\mathrm{const}$, or $v=\mathrm{const}*r$.  Thus,
iso-ram-pressure contours are just straight lines in this diagram.

\section{Summary and conclusions}
We presented spectroscopic observations of 96 disk galaxies in the Abell 901/902 multiple cluster system, which is located at a redshift of $z\sim 0.165$. Additionally, we compiled a comparison sample of 86 field galaxies at similar redshifts. The data were taken with the VLT instrument VIMOS. We carried out a redshift analysis and estimated dynamical parameters of the four subclusters, including their virial masses, virial radii, and velocity dispersions. For this we used biweight statistics \citep{beers90} and an interloper removal procedure suggested by \citet{perea90a}. We applied the Dressler-Shectman test \citep{ds88} to identify possible substructures in the cluster system. Furthermore, we extracted rotation curves from the spatially resolved emission lines and analysed distortions in the gaseous disk of the galaxies as well as HST/ACS images to quantify peculiarities of the stellar disk. We paid particular attention to dusty red galaxies as a possible intermediate stage in the transformation of field spirals to cluster S0 galaxies. The results of this work can be summarised as follows:\\
 \begin{itemize}
      \item[1.]
      The presence of substructures and non-Gaussian redshift distributions imply that the cluster system is dynamically young and in a non-virialised state. Accounting for interlopers, we derived the dynamical properties of the four subclusters.
	 \item[2.]
	 The fraction of kinematically distorted galaxies is $75\%$ higher in the cluster than in the field environment. This difference mainly stems from morphologically undistorted galaxies, indicating a cluster-specific interaction process which only affects the gas kinematics, not the stellar morphology. Ram-pressure stripping is therefore a prime candidate.
	 \item[3.]
	 We find two important galaxy populations:
	 \begin{itemize}
	 	\item Morphologically undistorted galaxies, which show high rotation-curve asymmetries at high rest-frame velocities and low cluster-centric radii, indicating that this group is strongly affected by ram-pressure stripping due to the interaction with the intracluster medium.
	 	\item Morphologically distorted galaxies, which are probably subject to tidal interactions. They show higher rotation-curve asymmetries at intermediate cluster-centric radii, and low rest-frame peculiar velocities.
     \end{itemize}
     \item[4.] Among the morphologically undistorted galaxies, the dusty red galaxies have particularly high rotation-curve asymmetries, suggesting an enhanced effect of ram pressure.
     \item[5.] The ratio between gas and stellar scale length is reduced for cluster galaxies compared to the field. This indicates that disks in the cluster environment lose part of their gas reservoir.
	\item[6.] The projected disk orientation of the cluster galaxies seems to be correlated with the gas scale length in the sense that:
	\begin{itemize}
		\item The ratio between the gas and stellar scale length steadily increases, going from face-on to edge-on infall.
		\item For more edge-on infall, the gas scale length of the leading edge is reduced compared to the gas scale length of the trailing edge of a galaxy.
	\end{itemize}
	Both findings suggest an influence of ram pressure, and are in agreement with numerical simulations. Furthermore, they indicate that the spiral galaxies in our sample are in the process of falling into, and on their first transit through, the cluster.
	\item[7.] Dusty red galaxies on average have bulge-to-total ratios  a factor of 2.5 higher than cluster blue cloud and field disk galaxies. This is further evidence that dusty red galaxies represent an intermediate stage in the transformation of field spirals to cluster lenticulars.
	\item[8.] Dusty red galaxies residing at the outskirts of the cluster have a higher rest-frame velocity. This might be indirect evidence for strong ram pressure producing this SED-type, as high velocities compensate the low ICM density at these radii.
	\item[9.] Galaxies with high velocities near the cluster centre are probably already quenched and are scarce in our sample due to our selection for galaxies with a star-forming SED.
\end{itemize}
We interpret the aforementioned results as indications that ram-pressure stripping is a key factor for galaxy transformations in the multiple cluster system A901/902, and that dusty red galaxies could be a transition state from spiral to S0 galaxies.

\begin{acknowledgements}
      We thank ESO for the support during the spectroscopic observations. We thank the anonymous referee for improving the quality of this work. For plotting, the 2D graphics environment \textit{Matplotlib} is used \citep{hunter07}. Asmus B\"{o}hm thanks the Austrian Science Fund (FWF) for funding (projects P19300-N16, P23946-N16).
\end{acknowledgements}

\bibliographystyle{aa}
\bibliography{benni}

\end{document}